\documentclass{aastex631}
\usepackage[T1]{fontenc}
\usepackage{amsmath}	
\usepackage{amssymb}	
\usepackage{bm}		

\begin{document}

\title{Radiation RMHD accretion flows around spinning AGNs: a comparative study of MAD and SANE state}



\correspondingauthor{Ramiz Aktar}
\email{ramizaktar@gmail.com}

\author[0000-0002-3672-6271]{Ramiz Aktar}
\affiliation{Department of Physics and Institute of Astronomy, National Tsing Hua University, 30013 Hsinchu, Taiwan}

\author[0000-0002-1473-9880]{Kuo-Chuan Pan}
\affiliation{Department of Physics and Institute of Astronomy, National Tsing Hua University, 30013 Hsinchu, Taiwan}
\affiliation{Center for Theory and Computation, National Tsing Hua University, Hsinchu 30013, Taiwan}
\affiliation{Physics Division, National Center for Theoretical Sciences, National Taiwan University, Taipei 10617, Taiwan}

\author{Toru Okuda}
\affiliation{Hakodate Campus, Hokkaido University of Education, Hachiman-Cho 1-2, Hakodate
040-8567, Japan}



\begin{abstract}

In our study, we examine a 2D radiation, relativistic, magnetohydrodynamics (Rad-RMHD) accretion flows around a spinning supermassive black hole. We begin by setting an initial equilibrium torus around the black hole, with an embedded initial magnetic field inside the torus. The strength of the initial magnetic field is determined by the plasma beta parameter, which is the ratio of the gas pressure to the magnetic pressure. In this paper, we perform a comparative study of the `magnetically arrested disc (MAD)' and `standard and normal evolution (SANE)' states. We observe that MAD state is possible for comparatively high initial magnetic field strength flow. Additionally, we also adopt a self-consistent two-temperature model to evaluate the luminosity and energy spectrum for our model. We observe that the total luminosity is mostly dominated by bremsstrahlung luminosity compared to the synchrotron luminosity due to the presence of highly dense torus. We also identify similar quasi-periodic oscillations (QPOs) for both MAD and SANE states based on power density spectrum analysis. Furthermore, our comparative study of the energy spectrum does not reveal any characteristic differences between MAD and SANE states. Lastly, we note that the MAD state is possible for both prograde and retrograde accretion flow.

\end{abstract}





\section{Introduction} \label{sec:intro}

Active galactic nuclei (AGNs) are generally powered by accretion onto supermassive black holes such as Sagittarius A* (Sgr A*) and Messier 87 (M87). Jets and outflows are also ubiquitously observed from AGNs. In this regard, highly relativistic jets have been observed in M87 \citep{Junor-etal-99, Cui-etal-23}. The M87 jet is well-collimated and propagates from pc to kpc scale. On the other hand, so far there is no clear evidence of jets for Sgr A*. However, there is some indication of the possibility of the existence of jet and outflow in Sgr A* \citep{Yusef-Zadeh-etal-20} though the jet velocities are not very high compared to the M87 jets. To understand the jet's ejection mechanism from black holes, one needs to investigate the accretion process in detail. In this regard, \citet{Shakura-Sunyaev73} first proposed the standard thin-disk model to investigate the accretion flows. They assumed that the disk is optically thick and supported by thermal pressure. Furthermore, the dissipation of gravitational binding energy is radiated away locally. Additionally, they utilized the famous "$\alpha$-disk" model to transport angular momentum in the disk. However, the standard thin-disk model has several problems both in terms of theoretical self-consistency and its ability to match observations \citep{Antonucci-13}. On the other hand, \citet{Balbus-Hawley91, Balbus-Hawley98} demonstrated that the magnetorotational instability (MRI) is an effective mechanism of angular momentum transport in the accretion ﬂows. MRI generates Maxwell and Reynolds stress to transport angular momentum outwards and causes inward mass accretion. Several MHD simulations have been carried out in the literature considering MRI turbulence in the flow \citep{Hawley-00, Machida-etal00, De-Villiers-etal03a, De-Villiers-etal03b, Kuwabara-etal05, Ohsuga-Mineshige11, Tchekhovskoy-etal11, Narayan-etal12, Igarashi-etal20, Chatterjee-Narayan-22, Dihingia-etal22, Hong-Xuan-etal23}.

Numerous studies have been conducted over the years to investigate the jets and outflows from black holes. The pioneering work by \citet{Blandford-Znajek77} (BZ) indicated that jet power is extracted from the rotational energy from large-scale magnetic ﬁelds around highly spinning black holes. Various simulation studies investigate the powerful jets around black holes based on BZ mechanism \citep{Tchekhovskoy-etal11, Tchekhovskoy-McKinney-12, Narayan-etal12, Dihingia-etal21, Chatterjee-Narayan-22, Hong-Xuan-etal23}. In this regard, \citet{Blandford-Payne82} (BP) explored mass outflow from the disk due to the magnetocentrifugal acceleration. Several other models have also been proposed in the literature to explain mass outflows or jets from black holes. These include the advection-dominated inﬂow-outﬂow solutions (ADIOS) \citep{Blandford-Begelman-99, Blandford-Begelman-04, Becker-etal-01, Xue-Wang-05, Begelman-12, Yuan-etal-12}, convection-dominated accretion ﬂow (CDAF) \citep{Igumenshchev-Abramowicz-99, Stone-etal-99, Igumenshchev-etal-00, Narayan-etal00, Quataert-Gruzinov-00, Igumenshchev-etal-03} and accretion shock-driven mass outﬂow model \citep{Chattopadhyay-Das07, Kumar-Chattopadhyay13, Das-etal14, Aktar-etal15, Aktar-etal17, Kim-etal-19, Okuda-etal19, Okuda-etal22, Okuda-etal23, Garain-Kim-23}, which have been investigated in the literature based on analytical as well as simulation studies.

In recent years, one of the interesting and important accretion states has been explored in highly magnetized accretion flows around the black hole, known as `magnetically arrested disc (MAD)' \citep{Narayan-etal-03}. The MAD state was first investigated by \citet{Igumenshchev-08}, who studied the Poynting jets using a simulation based on a pseudo-Newtonian approximation. Subsequently, \cite{Tchekhovskoy-etal11} showed that a hot accretion ﬂow exhibits MAD state, launching a very powerful jet (BZ) based on GRMHD simulation around a spinning black hole. They also explored the critical or threshold value of normalized magnetic flux accumulated at the horizon required for the flow to enter in the MAD state. In subsequent years, several independent simulation studies have investigated the magnetized flow, including MAD state and `standard and normal evolution (SANE)' \citep{McKinney-etal-12, Narayan-etal12, Dihingia-etal21, Mizuno-etal-21, Janiuk-James-22, Fromm-etal-22, Dhang-etal23, Hong-Xuan-etal23, Aktar-etal24, Zhang-etal-24}. In recent years, the observations of M87 and Sgr A* by the Event Horizon Telescope (EHT) collaboration have brought more attention to the MAD state \citep{Event-Horizon-etal-19, Event-Horizon-etal-21, Event-Horizon-etal-22}. It has been suggested that the accretion flow around these two black holes (AGNs) is in the MAD state, based on comparisons between radio images and post-processed GRMHD simulations. Additionally, the GRAVITY collaboration's high angular resolution near-infrared (NIR) observations of flares in Sgr A* have confirmed that their origin lies in magnetic flux eruption events, which are characteristic of the MAD state.

The radiatively inefficient accretion flows (RIAFs) model is good enough when the accretion rate is sufficiently very low. The RIAF model lies in the optically thin model for low accretion rates. In general, low luminosity AGNs (LLAGNs) are commonly observed and their luminosity lies $L \lesssim 0.1 L_{\rm Edd}$, where $L_{\rm Edd}$ is the Eddington luminosity. The LLAGNs involve thermally stable accretion onto black holes and are believed to form a geometrically thick, optically thin, radiatively inefficient accretion flow (RIAF) \citep{Narayan-etal95, Yuan-Narayan14}. However, the radiative process becomes crucial for the dynamics of the accretion flow when the accretion rates approach the Eddington accretion rates. But, the numerical modeling of radiation transport in black hole accretion is quite challenging due to the need to account for both optically thin and thick regions in the flow. In their pioneering work on global, non-relativistic, radiation hydrodynamics (RHD) simulations of super-Eddington accretion disks, \citet{Ohsunga-etal-05} implemented a radiation mechanism using the flux-limited diffusion technique. Following a similar approach, the radiation mechanism has been incorporated into the GRMHD simulation code by \citet{Sadowski-etal-13, Sadowski-etal-14} using the M1 closure scheme. Subsequently, various independent studies have explored radiation in black hole accretion disks, considering non-relativistic, relativistic, and general-relativistic flows \citep{Ohsuga-etal-09, Ohsuga-Mineshige11, McKinney-etal-14, Takahashi-etal-16, Okuda-etal22, Okuda-etal23}. In regions with sufficiently low-density accretion, the electron-proton Coulomb collision time is much longer than the dynamical timescale, leading to the development of a two-temperature structure in the flow, with protons and electrons having different temperatures. In recent years, several works have examined radiation transport mechanism based on a two-temperature model \citep{Ressler-etal-15, Ryan-etal-18, Chael-etal-19, Dihingia-etal23, Hong-Xuan-etal23, Okuda-etal23}.

In this paper, we investigate two-dimensional time-dependent, radiation, relativistic, magnetized accretion flow around spinning AGNs. We model the equilibrium MHD torus following \citet{Aktar-etal24} work. The space-time geometry around the black hole is modeled using effective-Kerr potential \citep{Dihingia-etal18b}. We perform a sufficiently high spatial resolution and long-term simulation to explore the properties of magnetized accretion flow. Here, we do not perform explicit two-temperature simulations to study accretion flow. However, we adopt a self-consistent two-temperature model following \cite{Okuda-etal23} to separately calculate electron and ion temperature in our model. We explicitly compare the MAD and SANE states based on our simulation. Additionally, in our model, we calculate the luminosity contribution from synchrotron and bremsstrahlung emission. Further, we attempt to compare the time evolution of the spectral energy distribution (SED) for MAD and SANE state following \citet{Manmoto-etal-97}. We also examine the effect of black hole spin on the highly magnetized accretion flow i.e., MAD state around black holes.

We organize the paper as follows. In Section \ref{simulation-scheme}, we present the description of the numerical model and governing equations. In  Section \ref{results}, we discuss the simulation results of our model in detail. Finally,  we draw the concluding remarks and summary in Section \ref{conclusion}.



\section{Simulation Setup}
\label{simulation-scheme}

We investigate two-dimensional (2D), relativistic and radiation magnetohydrodynamics (Rad-RMHD) simulations using freely available numerical simulation code PLUTO\footnote{\url{http://plutocode.ph.unito.it}} \citep{Migone-etal07, Fuksman-Mignone-19}. We use the unit system as $G = M_{\rm BH} = c =1$, where $G$, $M_{\rm BH}$ and $c$ are the gravitational constant, the mass of the black hole and the speed of light, respectively. Therefore, we measure distance, velocity, and time as $r_g = GM_{\rm BH}/c^2$, $c$ and $t_g = GM_{\rm BH}/c^3$, respectively. Here, we consider ideal special relativistic magnetohydrodynamics (RMHD) equations ignoring explicit resistivity in the flow.

\subsection{Governing Equations for Rad-RMHD}

Here, we present ideal RMHD governing equations for the interaction between matter
and electromagnetic (EM) fields in cylindrical coordinates $(r,\phi, z)$. The governing equations are as follows \citep{Fuksman-Mignone-19} \\
\begin{align}
& \label{governing_eq_1} \frac{\partial (\rho \gamma)}{\partial t} + \nabla \cdot (\rho \gamma \bm{v}) =0,\\
& \frac{\partial \varepsilon}{\partial t} + \nabla \cdot (\bm{m} - \rho \gamma \bm{v}) =G^0,\\
& \frac{\partial \bm{m}}{\partial t} + \nabla \cdot (\rho h \gamma^2 \bm{v} \bm{v} - \bm{B}\bm{B} - \bm{E} \bm{E}) + \nabla P_t = \bm{G}, \\
& \frac{\partial \bm{B} }{\partial t} + \nabla \times \bm{E} = 0,\\
& \frac{\partial E_{\rm rad}}{\partial t} + \nabla \cdot \bm{F}_{\rm rad} =-G^0,\\
& {\rm and}\\
& \label{governing_eq_7} \frac{\partial \bm{F}_{\rm rad}}{\partial t} + \nabla \cdot \bm{P}_{\rm rad} =-\bm{G},
\end{align}
where $\rho$, $\bm{v}$, $\bm{m}$, $h$, $\gamma$ and $\bm{B}$ are the mass density, fluid velocity, momentum density, specific enthalpy, Lorentz factor and mean magnetic field, respectively. $\bm{E} = -\bm{v} \times \bm{B}$ is the electric field. Moreover, $E_{\rm rad}$, $\bm{F}_{\rm rad}$ and $P_{\rm rad}$ are the radiation energy density, the radiation flux, and the pressure tensor as moments of the radiation field, respectively. Here, the total pressure, momentum density, and energy density of matter are given by
\begin{align}
& P_t = P_{\rm gas} + \frac{E^2 + B^2}{2},\\
& \bm{m} = \rho h \gamma^2 \bm{v} + \bm{E} \times \bm{B},\\
& \rm{and}\\
& \varepsilon = \rho h \gamma^2 - P_{\rm gas} - \rho \gamma + \frac{E^2 + B^2}{2}.
\end{align}
Also, here the radiation-matter interaction terms ($G^0, \bm{G}$) is given by
\begin{align}
(G^0, \bm{G})_{\rm comov} = \rho[\kappa (E_{\rm rad}-a_R T^4), (\kappa+\sigma) \bm{F}_{\rm rad}]_{\rm comov},
\end{align}
where all the fields are measured in the fluid’s comoving frame. Here $\kappa$ and $\sigma$ are the frequency-averaged absorption and scattering opacity, respectively. In this work, we adopt free-free absorption opacity as $\kappa = 1.7 \times 10^{-25} m_p^{-2} \rho T^{-7/2}~{\rm cm}^2~{\rm g}^{-1}$ and scattering opacity as $\sigma = 0.4~{\rm cm}^2~{\rm g}^{-1}$, respectively \citep{Igarashi-etal20}. Here, $T$ is the temperature of the fluid and $a_R$ is the radiation density constant. In this work, the radiation transfer equations are solved based on the gray approximation and imposing the M1 closure \citep{Fuksman-Mignone-19}.

\subsection{Closure relation of EoS and radiation fields}

In order to close the system of equations (\ref{governing_eq_1}-\ref{governing_eq_7}), it is required additional sets of relations. One of them is the equation of state (EoS) which provides closure between thermodynamical quantities. In this work, we consider constant-$\Gamma$ EoS, and is given as
\begin{align} \label{EoS_eq}
h = 1 + \frac{\Gamma}{\Gamma -1} \Theta,
\end{align}
where, $\Theta = \frac{P_{\rm gas}}{\rho}$ and $\Gamma$ is the adiabatic index. Further, another closure relation is needed for the radiation fields which relates to pressure tensor $P_{\rm rad}^{ij}$, $E_{\rm rad}$ and $\bm{F}_{\rm rad}$. The closure relation is as follows \citep{Fuksman-Mignone-19}
\begin{align}
& P_{\rm rad}^{ij} = D^{ij} E_{\rm rad}, \\
& D^{ij} = \frac{1-\xi}{2} \delta^{ij} + \frac{3 \xi -1}{2} n^i n^j,\\
& \xi = \frac{3 + 4 f^2}{5+2 \sqrt{4 - 3f^2}},
\end{align}
where, $\bm{n}=\bm{F}_{\rm rad}/|{\bm{F}_{\rm rad}}|$ and $f=|{\bm{F}_{\rm rad}}|/E_{\rm rad}$. Here $\delta^{ij}$ is the Kronecker delta. 

\subsection{Black hole gravitational potential}
\label{Kerr_pot}

In this work, the gravitational effect around a spinning black hole is modeled using effective Kerr potential proposed by \citet{Dihingia-etal18b}. The effective Kerr potential is as follows
 \begin{align} \label{DDMC18_pot}
 \Phi^{\rm eff}\left(r,z,a_k, \lambda\right)= \frac{1}{2} \ln \left(\frac{A (2 R - \Sigma) r^2 - 4 a_k^2 r ^4}{\Sigma \lambda \left(\Sigma \lambda  R^2 + 4 a_k r ^2 R - 2 \lambda  R^3\right)- A \Sigma r ^2 }\right),
 \end{align}
 where $R=\sqrt{r^2+z^2}$ = spherical radial distance, $\Delta=a_k^2+R^2-2 R$, $\Sigma=\frac{a_k^2 z^2}{R^2}+R^2$ and $A=\left(a_k^2+R^2\right)^2-\frac{a_k^2 r^2 \Delta}{R^2}$. Here, $\lambda$ and $a_k$ are specific flow angular momentum and black hole spin, respectively. The Keplerian angular momentum can be obtained from equation (\ref{DDMC18_pot}) in the equatorial plane $(z \rightarrow 0)$ as $\lambda_K = \sqrt{r^3 \frac{\partial \Phi^{\rm eff}}{\partial r}|_{\lambda \rightarrow 0}}$. The angular frequency is obtained as $\Omega = \lambda/r^2$. The effective Kerr potential is implemented in PLUTO code in a similar way as \citet{Aktar-etal24}. It is to be mentioned that \citet{Dihingia-etal18b} examined various comparative studies between GR and effective Kerr potential based on analytical studies. They showed an excellent close agreement of the transonic properties adopting effective Kerr potential in the semi-relativistic regime and GR around the Kerr black hole. Therefore, our numerical approach qualitatively retains the essential space-time features around black holes by implementing effective Kerr potential. Moreover, our simulation model could achieve higher spatial resolutions than GRMHD simulations for a given computing resource.

\subsection{Set up of initial equilibrium Torus}
\label{torus_set_up}

The initial equilibrium torus is constructed by adopting the Newtonian analog of relativistic tori as prescribed by \cite{Abramowicz-etal78}. In this work, we adopt the same formalism to set up the initial equilibrium torus as described by \citet{Aktar-etal24}. The density distribution of the torus is obtained by considering constant angular momentum flow \citep{Matsumoto-etal96, Hawley-00, Kuwabara-etal05, Aktar-etal24}
\begin{align}
\Phi^{\rm eff}\left(r,z, a_k, \lambda\right) + \frac{\Gamma}{\Gamma -1}\frac{P_{\rm gas}}{\rho}={\cal C},
\end{align}
where `$\mathcal{C}$' is the constant and it is calculated considering zero-gas pressure $(P_{\rm gas} \rightarrow 0)$ at $r = r_{\rm min}$ at the equatorial plane. Here, $r_{\rm min}$ is the inner edge of the torus. The density distribution inside the torus is obtained using the adiabatic equation of state $P_{\rm gas} = K \rho^{\Gamma}$ as
\begin{align}
 \rho=\left[\frac{\Gamma-1}{K\Gamma}\left({\cal C}-\Phi^{\rm eff}\left(r,z, a_k, \lambda\right)\right)\right]^{\frac{1}{\Gamma -1}},   
\end{align}
where $K$ is determined considering the density maximum condition $(\rho_{\rm max})$ at $r = r_{\rm max}$ at the equatorial plane and is given by
\begin{align}
 K=\frac{\Gamma - 1}{\Gamma}\left[\mathcal{C}-\Phi^{\rm eff}\left(r_{\rm max},0, a_k, \lambda\right)\right]\frac{1}{\rho_{\rm max}^{\Gamma-1}}.   
\end{align}
On the other hand, the density distribution outside the torus can be obtained by assuming hydrostatic equilibrium condition \citep{Matsumoto-etal96, Kuwabara-etal05, Aktar-etal24}. Therefore, the density outside the torus is assumed to be the isothermal, nonrotating, high-temperature halo, and is given by \citep{Manmoto-etal-97, Kuwabara-etal05, Igarashi-etal20}
\begin{align} \label{eqn_density_outside}
\rho={\eta}\rho_{\rm max}\exp\left[\left(\Phi^{\rm eff}\left(r_{\rm max},z, a_k, \lambda \rightarrow 0\right)-\Phi^{\rm eff}\left(r,z, a_k, \lambda \rightarrow 0\right)\right){\cal H}\right],    
\end{align}
where, $\cal{H}$ = $1/c_s^2$ and $c_s = \Gamma P_{\rm gas}/\rho$ is the sound speed. Here, we consider $\eta = 10^{-4}$ and $\cal{H}$=2 for the purpose of representation \citep{Aktar-etal24}.

\begin{table*}
\centering
\caption{Simulation models 
\label{Table-1}}
 \begin{tabular}{@{}c c c c c c c  } 
 \hline
 Model &   $r_{\rm min}$ &   $r_{\rm max}$  & $\rho_{\rm max}$  & $E_{\rm rad}$  &  $\beta_0$  & $a_k$ \\ 
       &     ($r_g$)     &   ($r_g$)   &   (g cm$^{-3}$)   & (erg  cm$^{-3}$)    &      &        \\
 \hline  
 $\beta 10$  &  40  &  60  &  $1 \times 10^{-10}$ & $5 \times 10^{-10}$  &   10  & 0.98  \\
 $\beta 25$  &  ''   &  ''   &   ''    &   ''   &   25  &  '' \\
 $\beta 50$  &  ''   &  ''   &   ''    &   ''   &   50  &  '' \\
 $\beta 100$ &  ''   &  ''  &    ''   &    ''  &  100  &   ''  \\ \\

 $a99$  &  ''    &   ''   &    ''      & '' &   10  & 0.99  \\
 $a50$  &  ''    &   ''   &    ''     & ''  & ''  & 0.50  \\
 $a00$  &  ''    &   ''   &    ''     & '' & '' & 0.00  \\
 $am99$ &  ''    &   ''   &    ''     & '' & '' & -0.99  \\
 \hline

 \end{tabular}
\end{table*}


\begin{figure*}
	\begin{center}
        \includegraphics[width=0.48\textwidth]{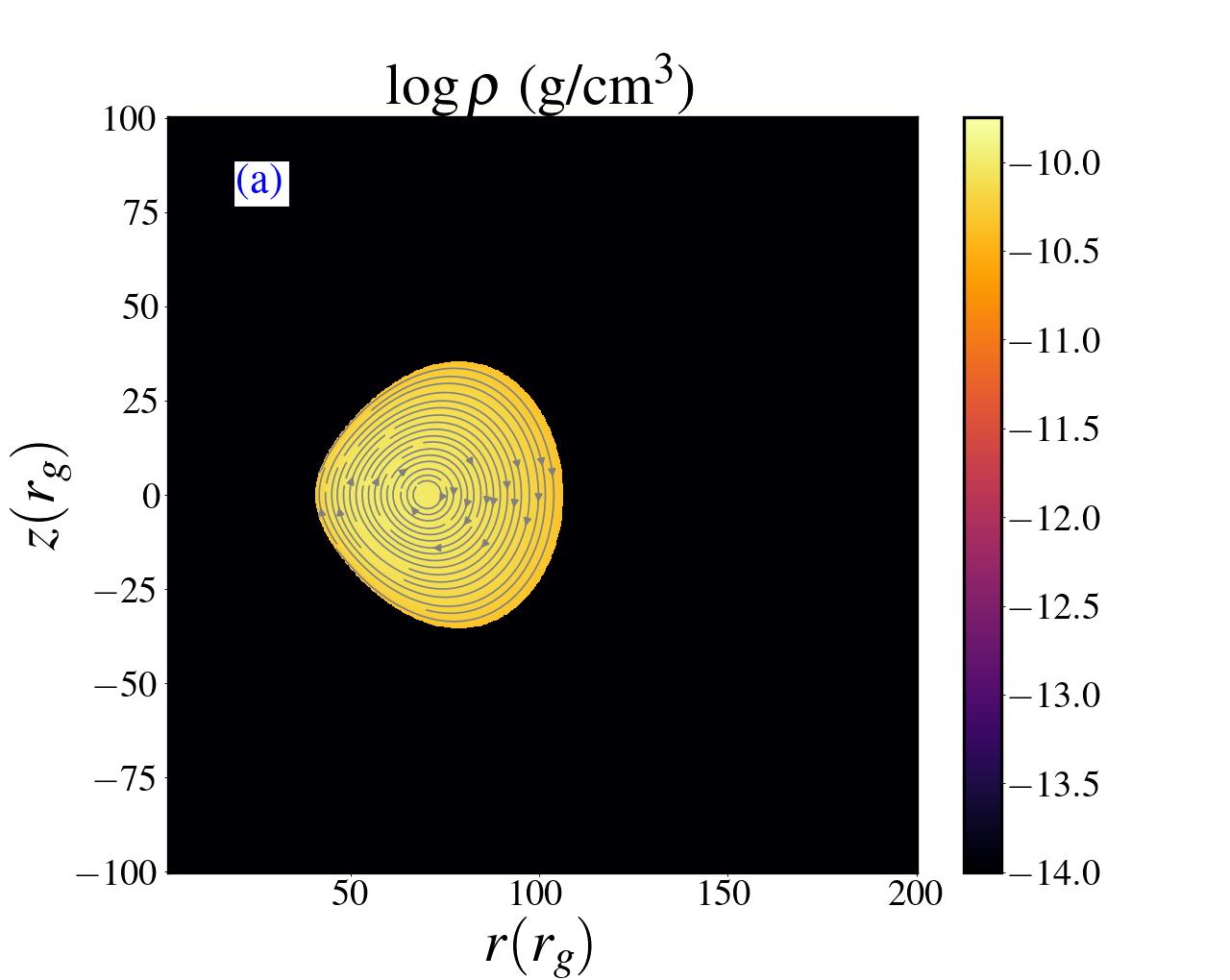} 
        \includegraphics[width=0.48\textwidth]{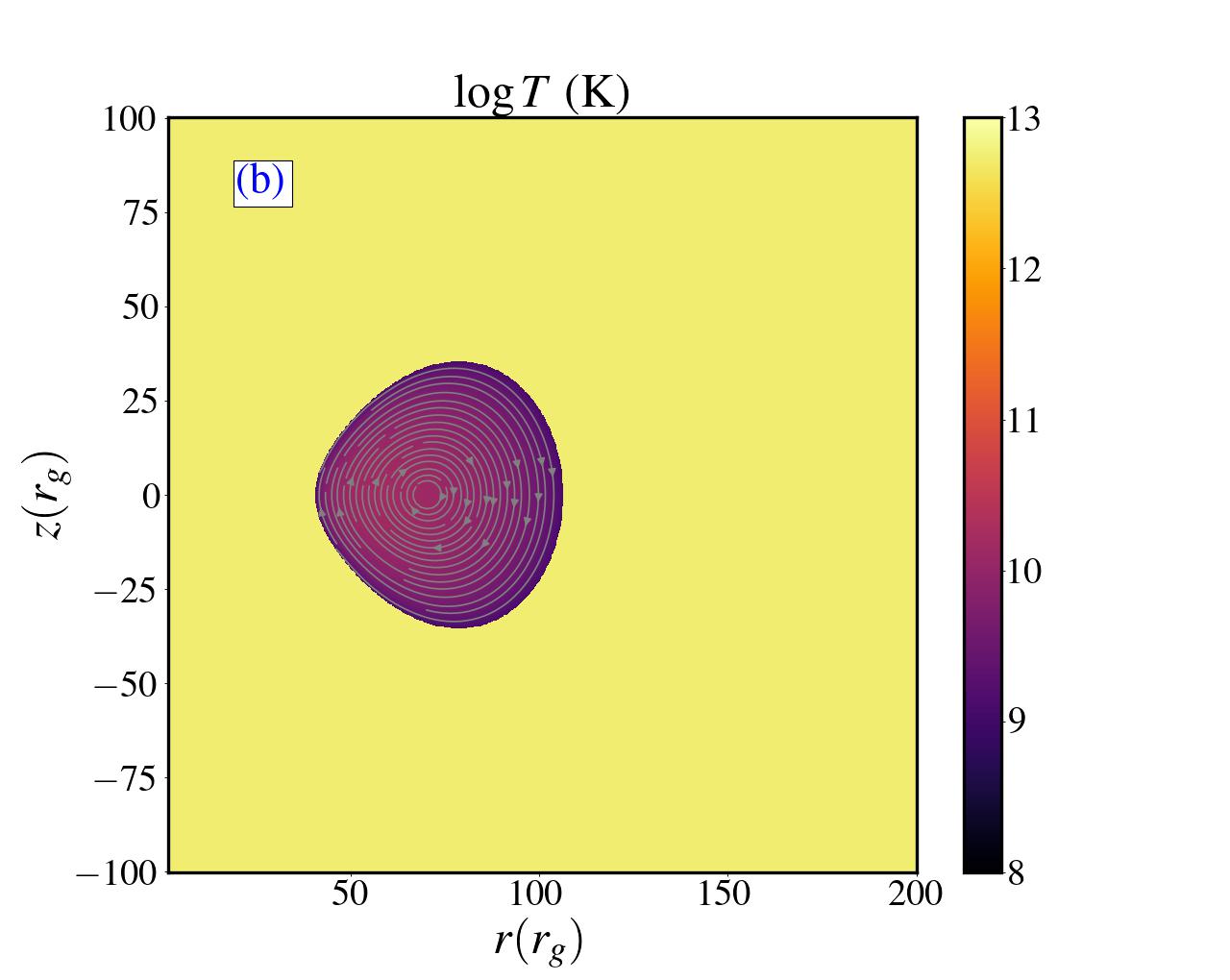} 
	\end{center}
	\caption{Initial equilibrium torus of $(a)$: density ($\log \rho $) and $(b)$: temperature ($\log T $) for $a_k = 0.98$ and $\beta_0 = 10$ (Model $\beta 10$). The grey lines represent magnetic field lines.}
	\label{Figure_01}
\end{figure*}

\begin{figure}
	\begin{center}
        \includegraphics[width=0.49\textwidth]{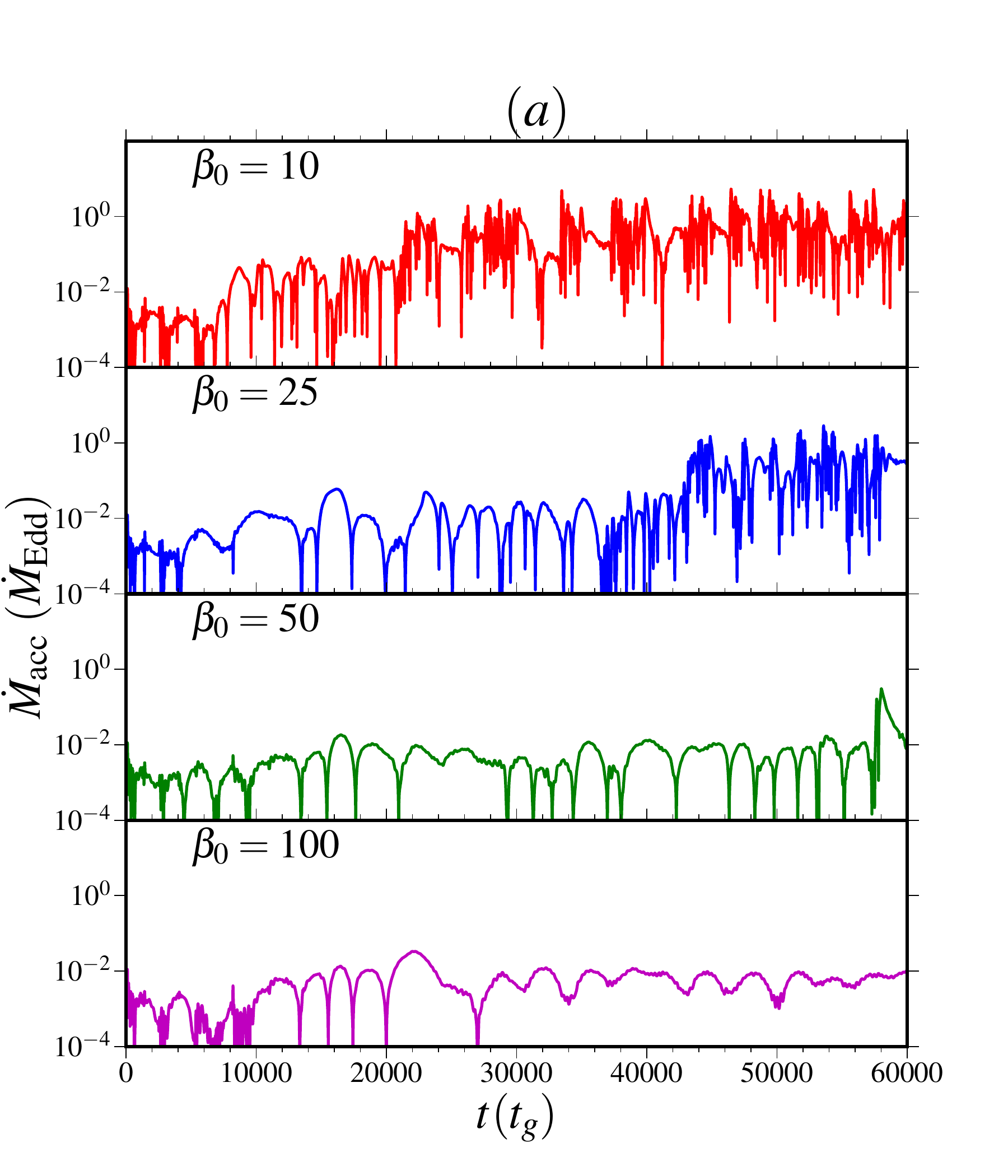} 
		\includegraphics[width=0.49\textwidth]{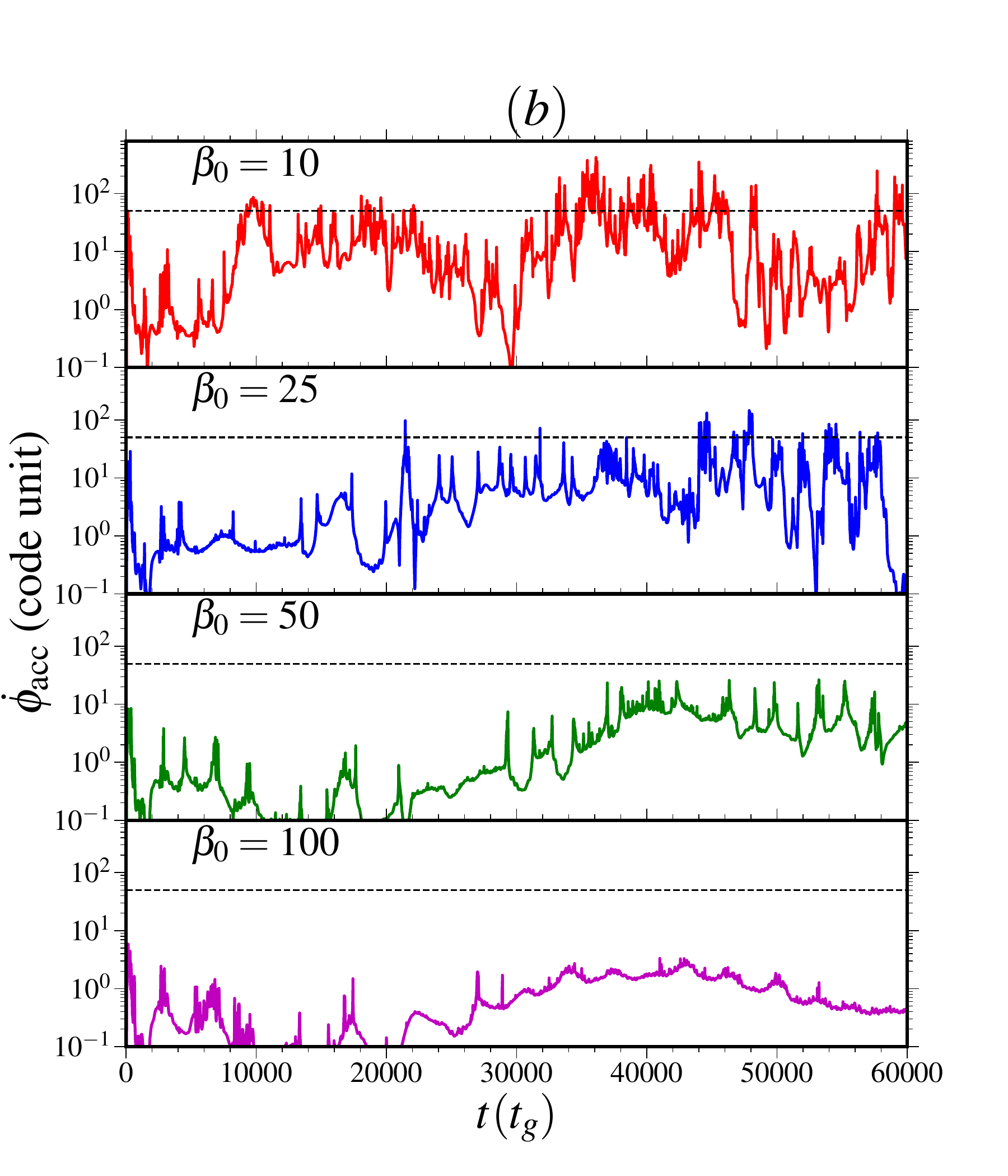} 
	\end{center}
	\caption{Temporal evolution of $(a)$: mass accretion rate $\dot{M}_{\rm acc} (\dot{M}_{\rm Edd})$
    and $(b)$: normalized magnetic ﬂux $\dot{\phi}_{\rm acc}$ in code units accumulated at the 
    black hole inner boundary with the simulation time for different initial plasma-$\beta$ parameter $\beta_0$. Dashed horizontal lines are for $\dot{\phi}_{\rm acc}$ = 50. Here, we consider the initial plasma-$\beta$ parameter as $\beta_0$ = 10, 25, 50 and 100. We also fix the black hole spin $a_k = 0.98$. See the text for details.}
	\label{Figure_1}
\end{figure}

\begin{figure*}
	\begin{center}
        \includegraphics[width=0.25\textwidth]{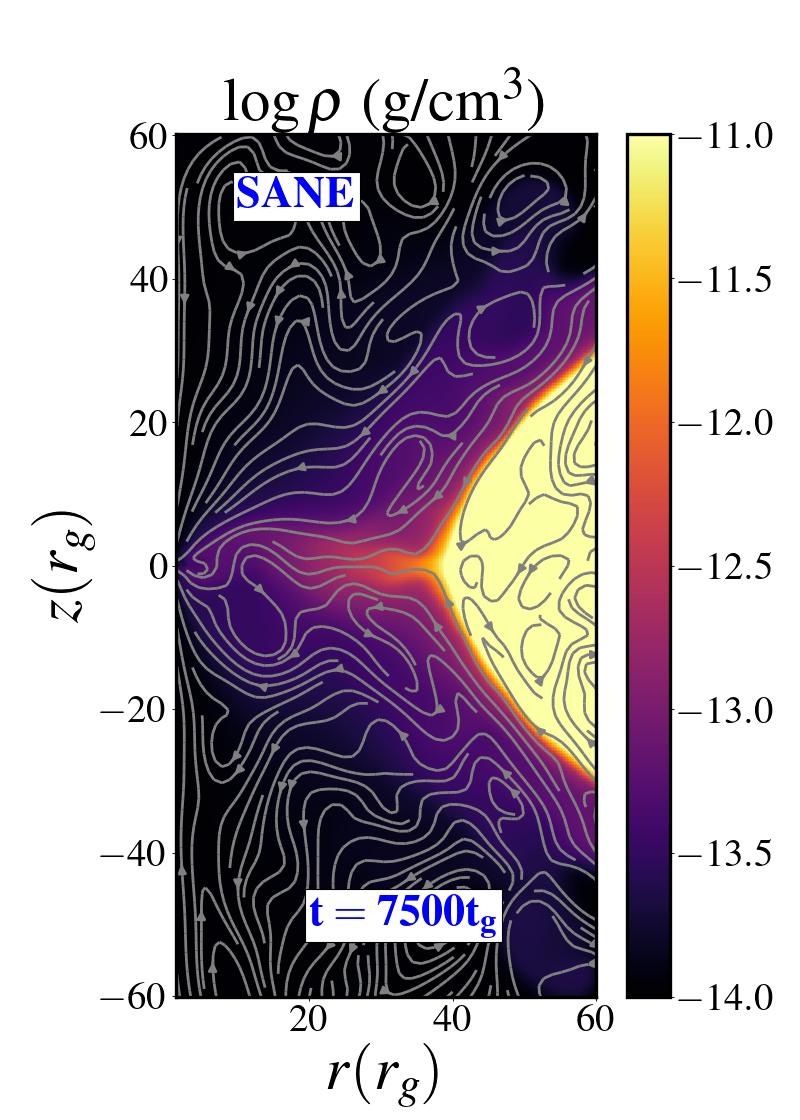} 
        \hskip -2mm
        \includegraphics[width=0.25\textwidth]{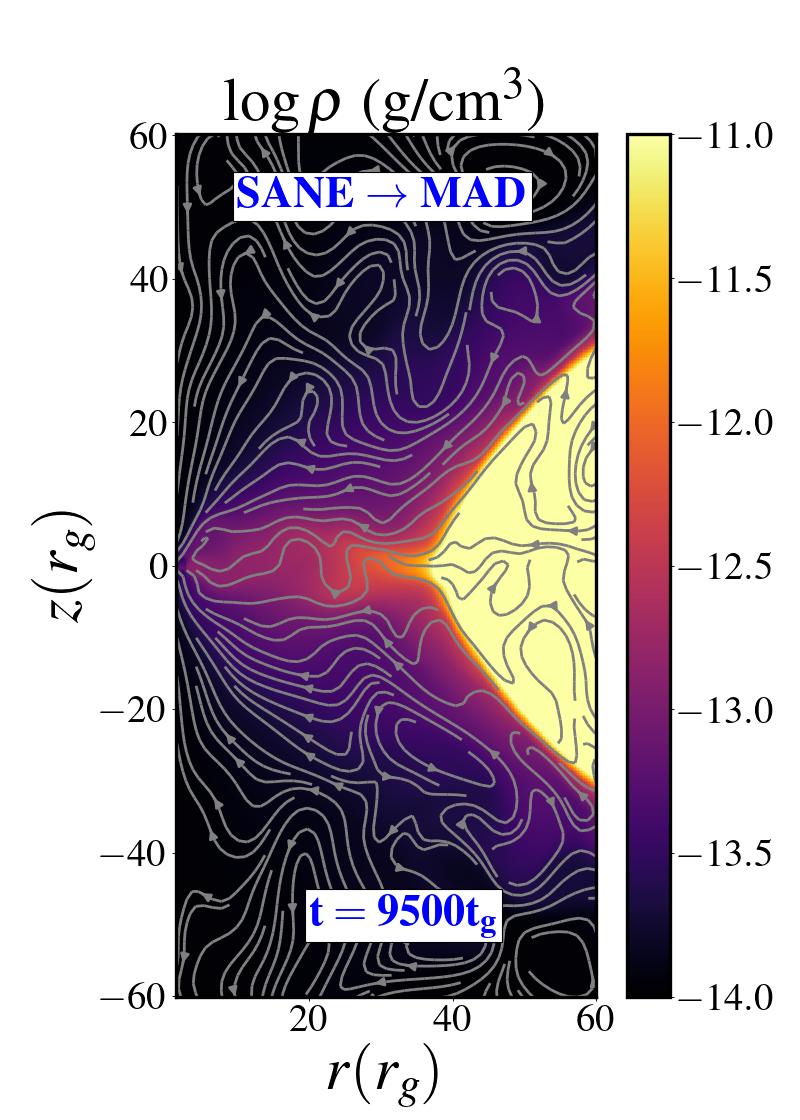} 
        \hskip -2mm
		\includegraphics[width=0.25\textwidth]{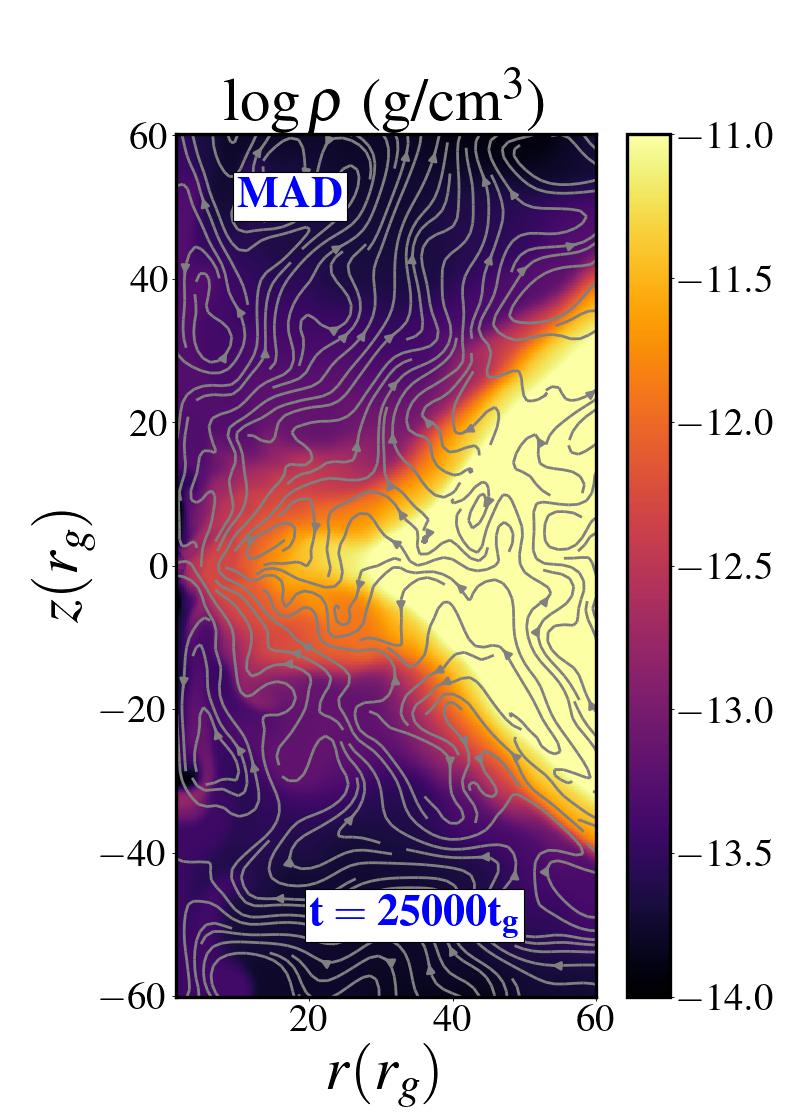} 
        \hskip -2mm
        \includegraphics[width=0.25\textwidth]{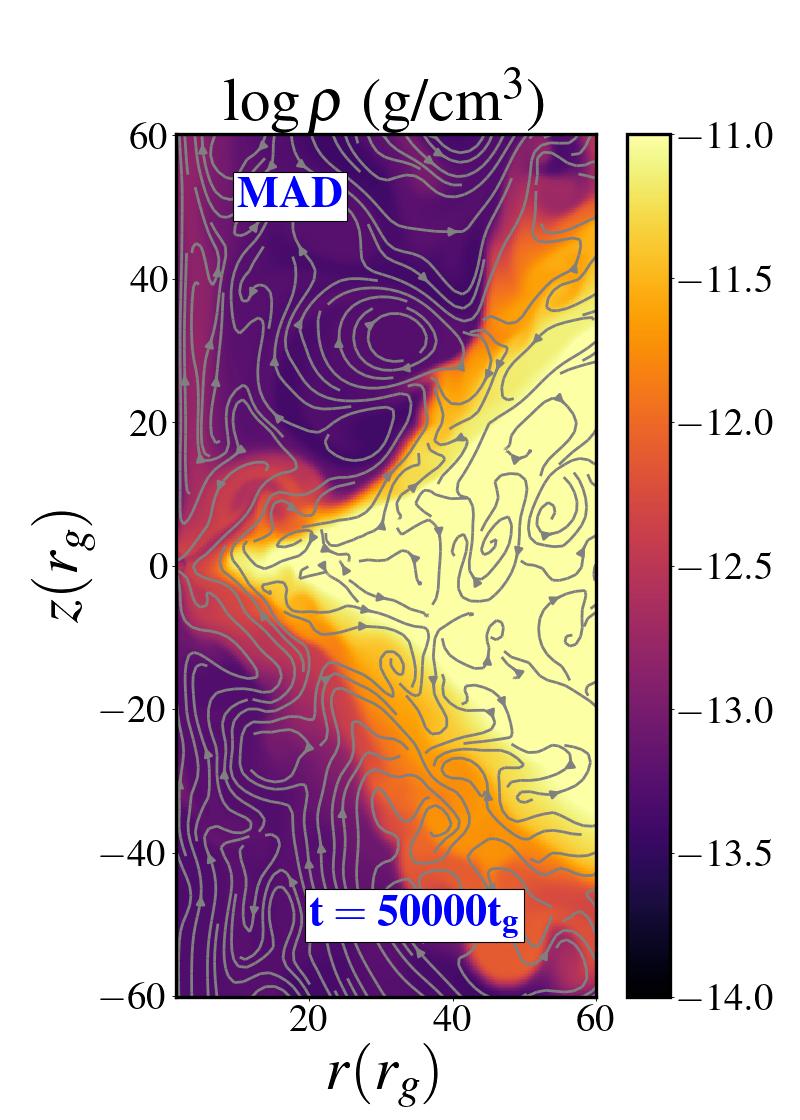} 

        \includegraphics[width=0.25\textwidth]{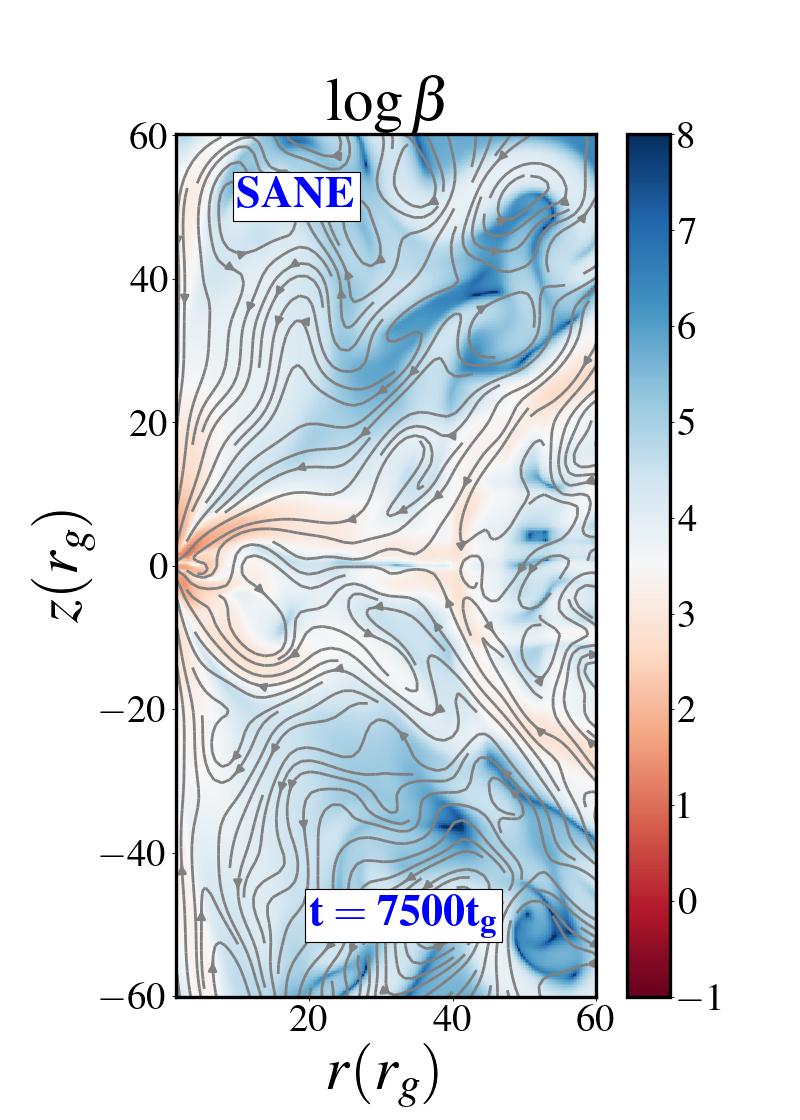} 
        \hskip -2mm
        \includegraphics[width=0.25\textwidth]{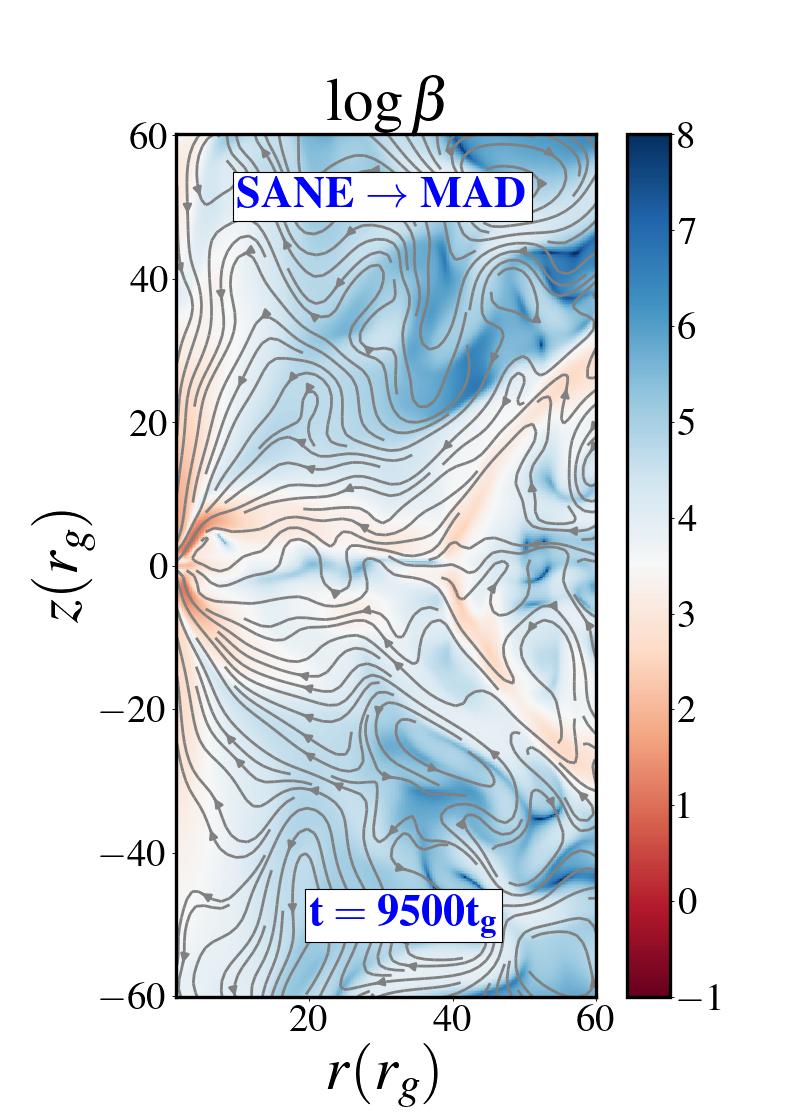} 
        \hskip -2mm
		\includegraphics[width=0.25\textwidth]{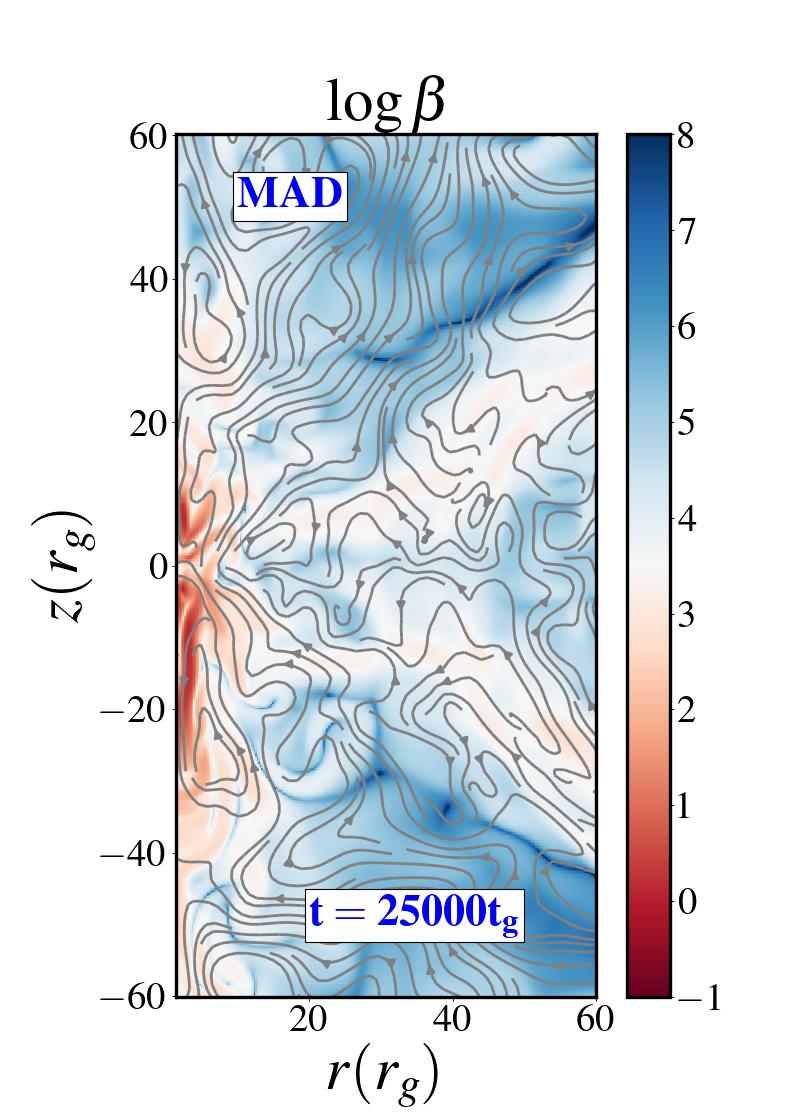} 
        \hskip -2mm
        \includegraphics[width=0.25\textwidth]{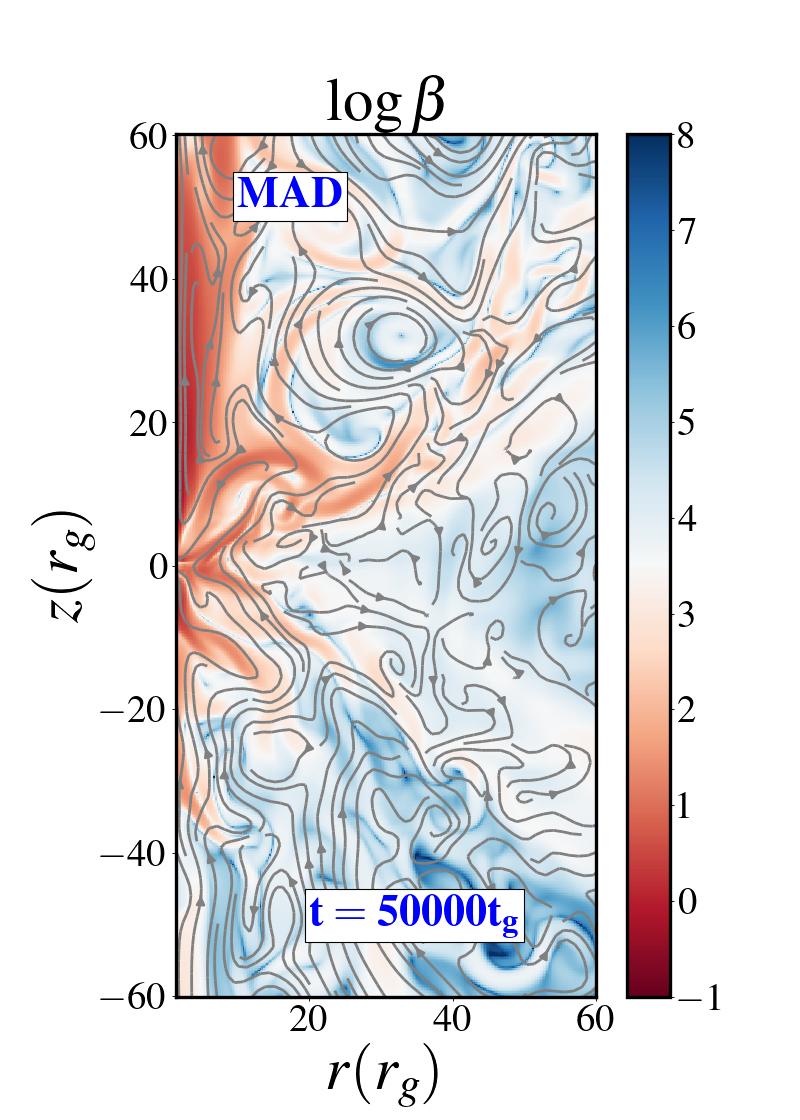} 
	\end{center}
	\caption{Distribution of gas density $(\rho)$ (upper panel) and plasma-$\beta$ $(\beta)$ (lower panel) for MAD state with the time evolution. Here, we consider the MAD sate  for $\beta_0 = 10$. The grey lines represent the velocity field lines. See the text for details.}
	\label{Figure_2}
\end{figure*}

\begin{figure*}
	\begin{center}
        \includegraphics[width=0.25\textwidth]{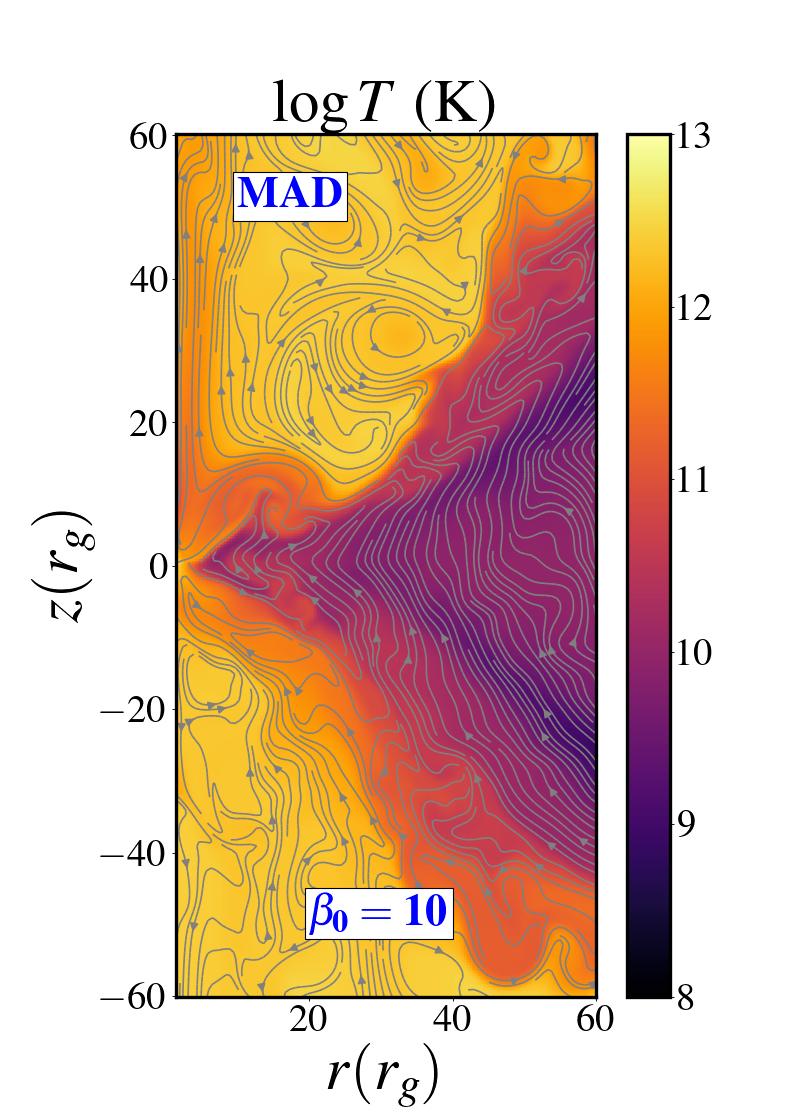} 
        \hskip -2mm
		\includegraphics[width=0.25\textwidth]{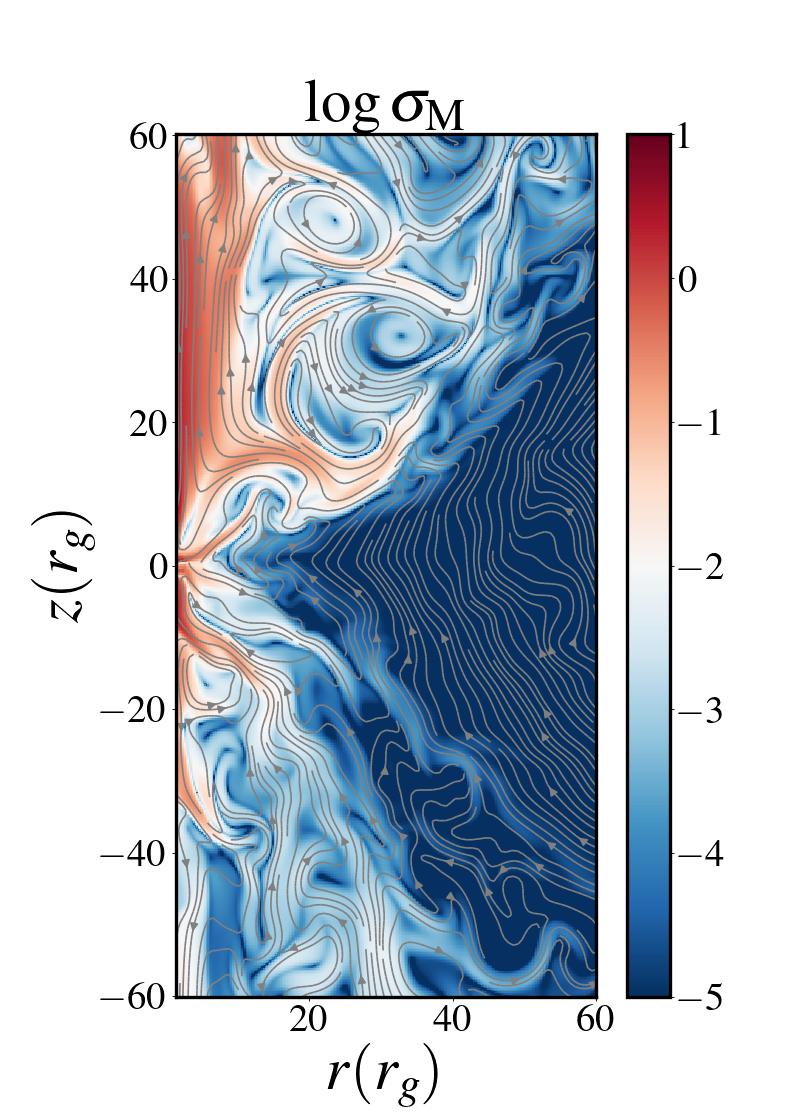}
        \hskip -2mm
        \includegraphics[width=0.25\textwidth]{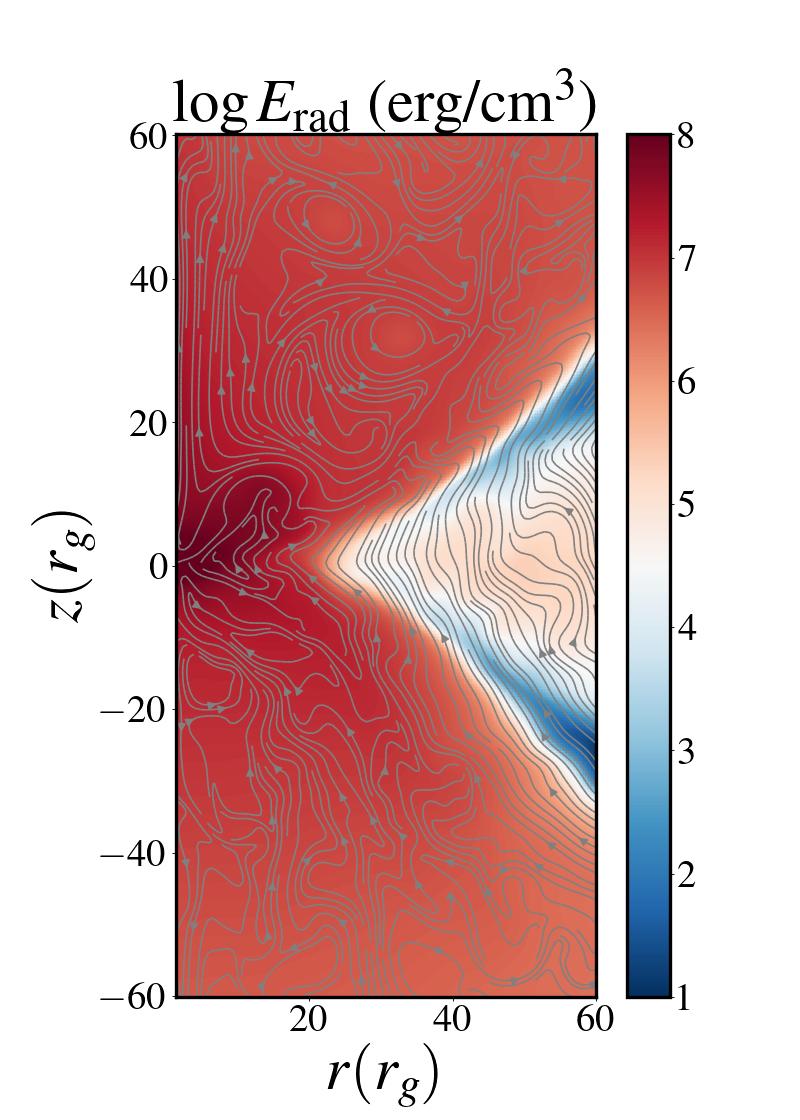} 
        \hskip -2mm
        \includegraphics[width=0.25\textwidth]{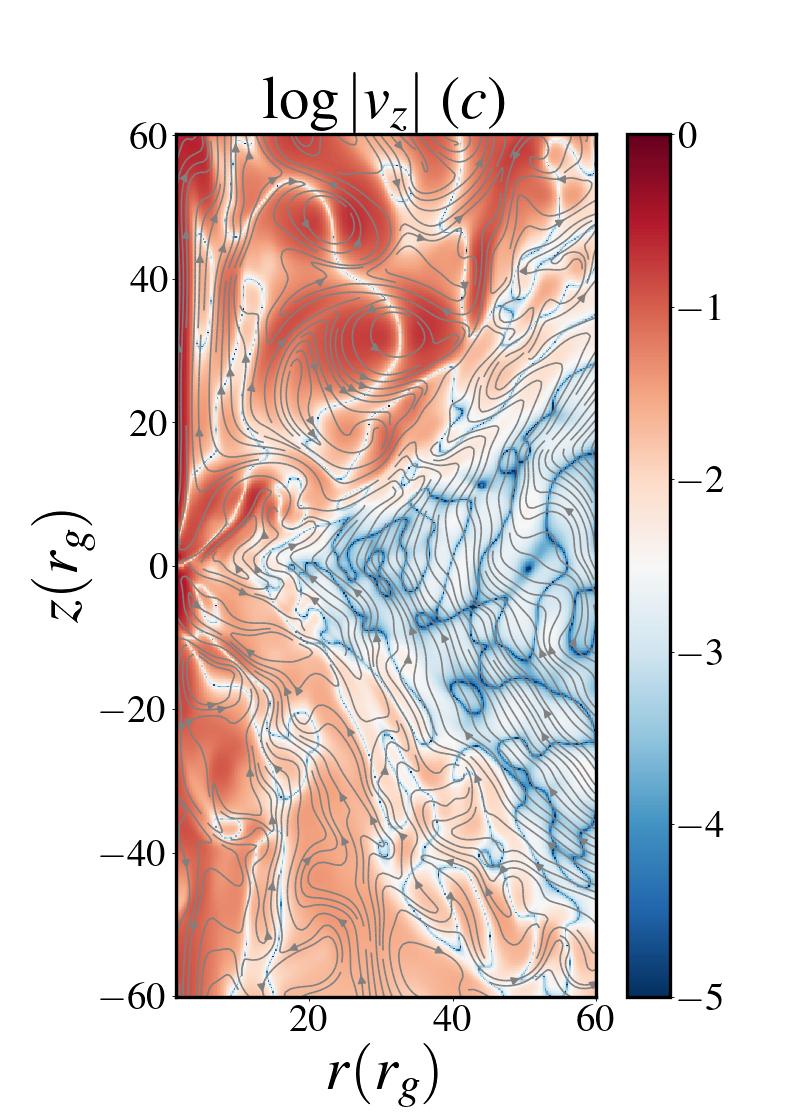} 
        
        \includegraphics[width=0.25\textwidth]{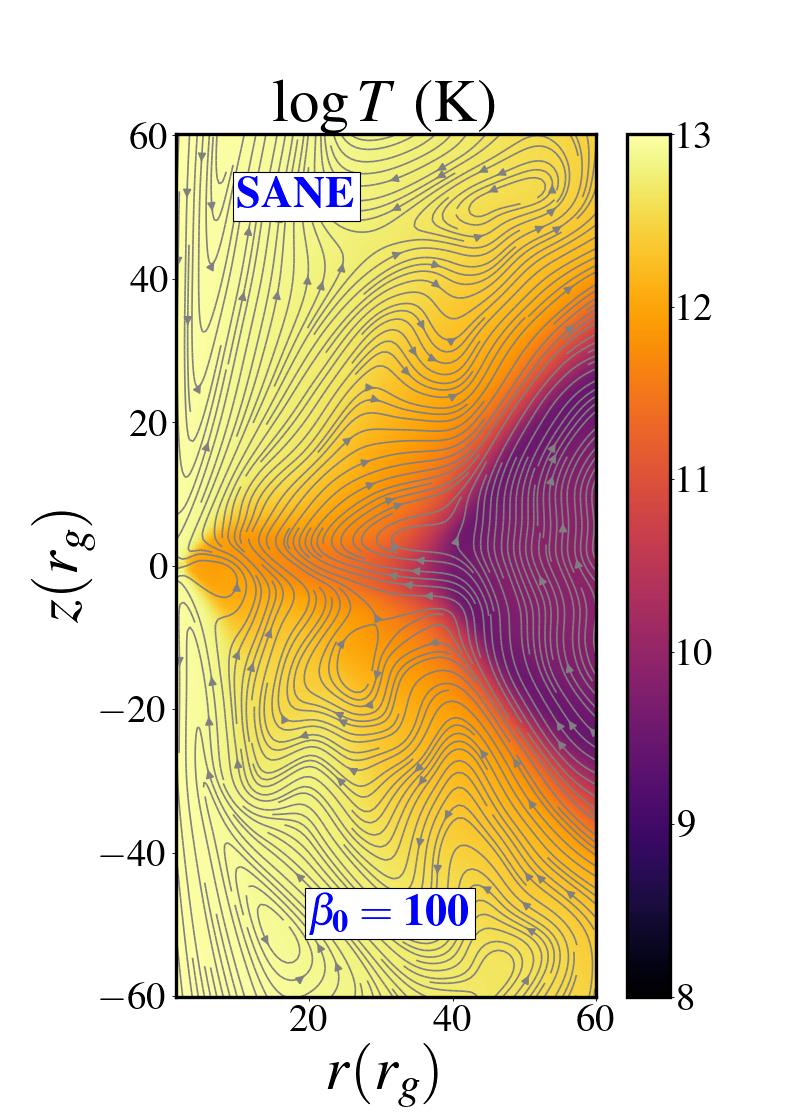} 
        \hskip -2mm
		\includegraphics[width=0.25\textwidth]{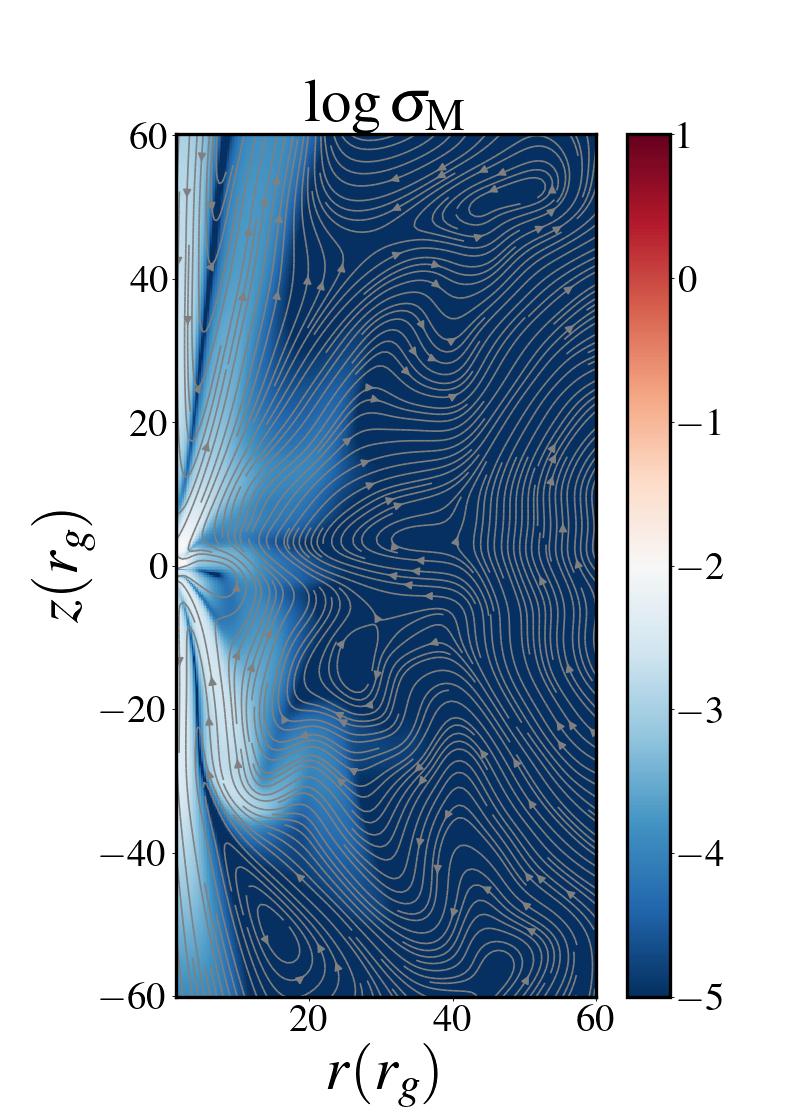} 
        \hskip -2mm
        \includegraphics[width=0.25\textwidth]{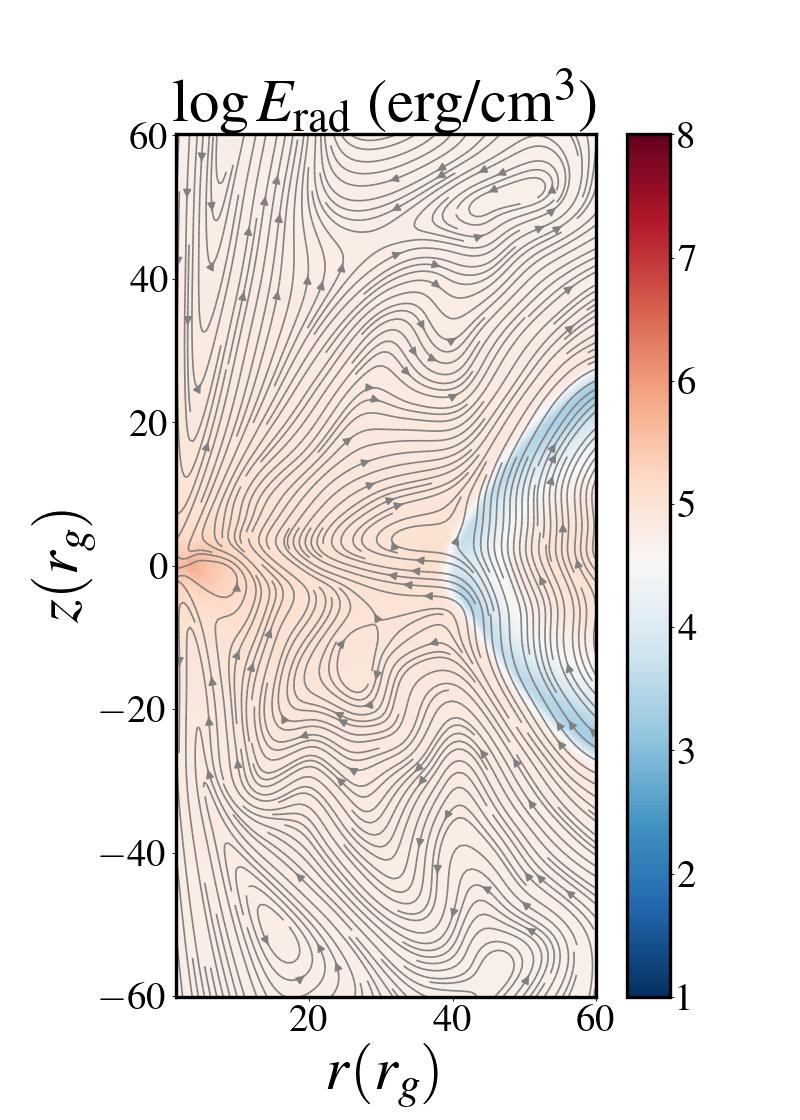} 
        \hskip -2mm
        \includegraphics[width=0.25\textwidth]{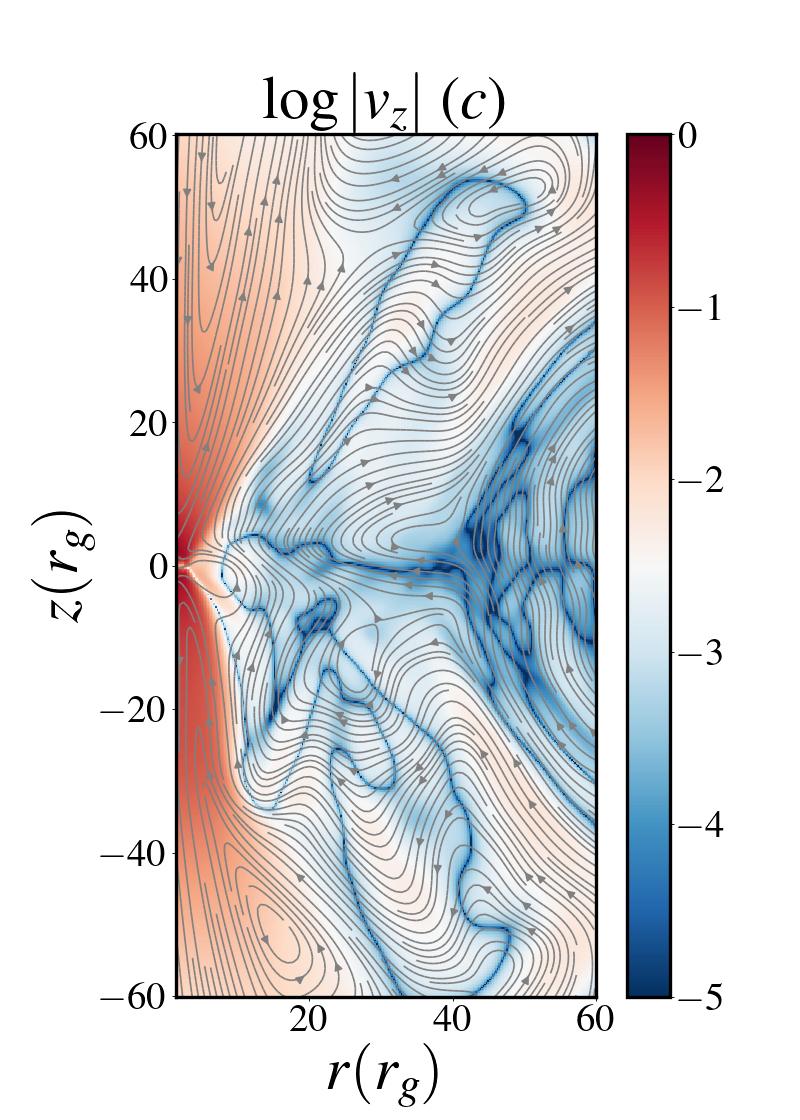} 
	\end{center}
	\caption{Comparison of temperature $(T)$, magnetization parameter $(\sigma_{\rm M})$, radiation energy density ($E_{\rm rad}$) and magnitude of vertical velocity $(|v_z|)$ in the ($r-z$) plane, respectively. Upper rows and lower rows are for MAD sate ($\beta_0 = 10$) and SANE state ($\beta_0 =100$), respectively at time $t = 50000 t_g$. The grey lines represent the magnetic field lines. See the text for details.}
	\label{Figure_3}
\end{figure*}

\subsection{Initial magnetic field configuration in the torus}
\label{Mag_config}

In this paper, we initialize a poloidal magnetic field by applying a purely toroidal component of vector potential \citep{Hawley-etal02}. The expression of the toroidal component of vector potential is as follows \citep{Hawley-etal02, Aktar-etal24}
\begin{align}
 A_\phi = B_0 [\rho(r,z) - \rho_{\rm min}],   
\end{align}
where, $B_0$ is the normalized initial magnetic field strength and $\rho_{\rm min}$ is the minimum density in the torus. Here the initial magnetic field strength is parameterized by the ratio of the gas pressure to the magnetic pressure known as plasma-$\beta$ parameter $\beta_0 =  \frac {2 P_{\rm gas}}{B_0^2}$. Also, we define the magnetization parameter as $\sigma_{\rm M} = \frac{B^2}{\rho}$ \citep{Dihingia-etal21, Dihingia-etal22, Dhang-etal23, Curd-Narayan23, Aktar-etal24}. The magnetization parameter essentially represents the ratio of magnetic energy to the rest mass energy in the flow.

\subsection{Initial and boundary conditions}

In this work, to simulate radiation relativistic MHD accretion flows, we adopt PLUTO simulation code \citep{Fuksman-Mignone-19}. Here, we consider the computational domain for our simulation model in the radial and vertical direction as $1.5 r_g \leq r \leq 200 r_g$ and $-100 r_g \leq z \leq 100 r_g$, respectively. We adopt a comparatively high resolution of grid sizes as $(n_r, n_z) = (920, 920)$ and uniform grid spacing in both radial and azimuthal directions. Here, we use the HLL Riemann solver, second-order-in-space linear interpolation, and the second-order-in-time Runge-Kutta algorithm. We impose the hyperbolic divergence cleaning method for solving the induction equation $\nabla \cdot \bm{B} = 0$ \citep{Migone-etal07}. The inner boundary is set to be absorbing boundary conditions around black hole \citep{Okuda-etal19, Okuda-etal22, Okuda-etal23, Aktar-etal24}. Here, we employ the inner boundary at $R_{\rm in} = 2.5 r_g$ which acts as the event horizon location for our model. Further, we set axisymmetric boundary conditions at the origin, and all the rest are set to outflow boundary conditions \citep{Aktar-etal24}.

To set up the initial equilibrium torus, we consider the inner edge of the torus at $r_{\rm min} = 40 r_g$ and the maximum pressure surface at $r_{\rm max} = 60 r_g$. We set the minimum density of the torus $\rho_{\rm min} = 0.5 \rho_{\rm max}$, where $\rho_{\rm max}$ is the maximum density of the torus. We choose $\rho_{\rm max} = \rho_0$ for our simulation, where $\rho_0$ is the reference density. It is widely accepted in the literature that there are consistent findings regarding the sizes of the torus for the MAD and SANE states in GRMHD simulations \citep{Wong-etal-21, Fromm-etal-22}. For the MAD state, the inner edge ($r_{\rm min}$) and maximum pressure location ($r_{\rm max}$) are typically situated at approximately $r_{\rm min} \approx 20 r_g$ and $r_{\rm max} \approx 40 r_g$ respectively. Conversely, for the SANE state, $r_{\rm min} \approx 6 r_g$ and $r_{\rm max} \approx 12 r_g$. Additionally, the initial magnetic field is set differently for the MAD and SANE states \citep{Wong-etal-21, Fromm-etal-22, Dihingia-etal23}. However, the primary focus of this study is to analyze the impact of initial magnetic fields on accretion states (MAD and SANE) when considering the same initial magnetic field configuration (see sub-section \ref{Mag_config}), to compare luminosity, and to investigate the spectral evolution of black holes without altering the geometrical configuration of the torus. For solving radiation fields, we need to supply additionally the radiation energy density $E_{\rm rad}$. In this work, we fix the reference density $\rho_0 = 10^{-10}$ g cm$^{-3}$ and the radiation energy density $E_{\rm rad} = 5 \times 10^{-10}$ erg cm$^{-3}$. Also, the mass of the black hole is chosen as $10^8 M_{\odot}$, and the constant adiabatic index is $\Gamma = 4/3$. The conversion relation between the code unit and C.G.S. unit system is summarized in \citet{Aktar-etal24}.

To avoid the situation of encountering negative density and pressure in supersonic and highly magnetized flows, we fix the floor values of density and pressure as $\rho_{\rm floor} = 10^{-6}$ and $P_{\rm floor} = 10^{-8}$, respectively. Along with that, we also turn on two flags (i) SHOCK  FLATTENING to MULTID and (ii) FAILSAFE to YES in PLUTO module \citep{Aktar-etal24}. We also impose the magnetization condition as $\sigma_{\rm M}< 4 \pi v_p^2$ for all the radii for the entire computational domain \citep{Proga-Begelman-03, Aktar-etal24}. Here, $v_p$ is the poloidal fluid velocity. Therefore, the floor value of magnetization is $\sigma_{\rm M}^{\rm floor}< 4 \pi v_p^2$ in our simulation model.

\section{Simulation Results}
\label{results}

To perform the simulation, we first set up the initial equilibrium torus around the black hole. We supply the specific Keplerian angular momentum $(\lambda_{\rm K})$ and spin of the black hole $(a_k)$ to form the equilibrium torus. The initial magnetic field is embedded within the torus (see Fig. 1 of \citet{Aktar-etal24}). We illustrate the density distribution ($\rho$) and temperature ($T$) of the initial equilibrium torus at time $t = 0$, as depicted in Fig. \ref{Figure_01}a and Fig. \ref{Figure_01}b, respectively. The outer edge of the torus is well within the computational domain, so the outflow boundary conditions at the outer edge of the computational domain do not impact the evolution of the torus. The initial temperature of the atmosphere outside the torus is hot and rarefied, as depicted in Fig. \ref{Figure_01}b. With time, MRI grows in the disk, and MRI transports angular momentum outwards \citep{Balbus-Hawley91}. Consequently, the accretion gradually increases towards the event horizon of the black hole. Therefore, accretion flow becomes more turbulent depending on the initial magnetic field and the gaseous matter spreads vertically. At times, the gaseous matter escapes as mass outflows due to the enhanced magnetic pressure in the inner disk \citep{Machida-etal00, Hawley-Krolik01, Hawley-Balbus02, De-Villiers-etal03a, Aktar-etal24}. In the first section of the paper, we investigate the effect of the magnetic field by varying initial plasma-$\beta$ ($\beta_0$) parameters for fixed black hole spin parameter ($a_k$). In the last section of the paper, we examine the effect of black hole spin by varying the black hole spin for the fixed magnetic field. All the simulation models are tabulated in Table \ref{Table-1}.  

In MHD flows, it is necessary to resolve the fastest-growing MRI mode in the simulation model. In general, the quality factor to resolve the MRI is defined as $Q_r = 2 \pi V_{Ar}/\Omega \Delta r$ and $Q_z = 2 \pi V_{Az}/\Omega \Delta z$ \citep{Hawley-etal11, Hawley-etal13}. Here, $V_{Ar}$ and $V_{Az}$ are the radial and vertical components of Alfv\'en speed, respectively. Also, $\Delta r$ and $\Delta z$ are grid sizes in radial and vertical directions, respectively. We calculate the average values of $Q_r$ and $Q_z$ by taking the time average over the interval $40000t_g < t< 50000t_g$ and the vertical space average over $-5 r_g <r < 5r_g$. Our simulation results show that the average values of $Q_r$ and $Q_z$ are generally greater than 15 across the entire computational domain. This indicates that our simulation model can effectively resolve the MRI \citep{Aktar-etal24}.

\subsection{Investigation of magnetic state}
\label{magnetic_state}

To investigate the magnetic state of our model, we investigate magnetic flux accumulated at the horizon. In general, we deﬁne dimensionless normalized magnetic ﬂux threading to the black hole horizon $(\dot{\phi}_{\rm acc})$ known as the MAD parameter. Here, the MAD parameter is defined as \cite{Tchekhovskoy-etal11, Narayan-etal12, Dihingia-etal21, Dhang-etal23, Aktar-etal24}
\begin{align}\label{mag_flux_acc_eqn}
\dot{\phi}_{\rm acc} = \frac{\sqrt{4 \pi}}{2 \sqrt{\dot{M}_{\rm acc}}}  \int{|B_r|_{R = R_{\rm in}} dz},
\end{align}
where $R_{\rm in}$ is the inner boundary representing the horizon for our model. $B_r$ is the radial component of the magnetic field. In our model, we define normalized magnetic flux in Gaussian units by multiplying the factor of $\sqrt{4 \pi}$ \citep{Narayan-etal-22}. Here, $\dot{M}_{\rm acc}$ is the mass accretion rate and given as
\begin{align}\label{mass_acc_eqn}
\dot{M}_{\rm acc} = - 2\pi \int{\rho (r,z) r v_r dz}, 
\end{align}
where the negative sign refers to the inward direction of the accretion flow. In Fig. \ref{Figure_1}a, we represent the temporal variation of mass accretion rate in Eddington units. Every panel represents different initial plasma-$\beta$ parameters such as $\beta_0 = 10, 25, 50$, and 100, respectively (see Table \ref{Table-1}). Here, we fix the spin of the black hole at $a_k =0.98$. We observe that the mass accretion rate saturates close to Eddington mass accretion rate ($\dot{M}_{\rm acc} \sim \dot{M}_{\rm Edd}$) for $\beta_0 = 10$ and 25. However, the mass accretion rate remains in the sub-Eddington limit ($\dot{M}_{\rm acc}<< \dot{M}_{\rm Edd}$) for $\beta_0 = 50$ and 100. One of the good indicators for examining the magnetic state is to investigate the value of the normalized magnetic flux entering through the horizon. Therefore, we also present the corresponding normalized magnetic flux $(\dot{\phi}_{\rm acc})$ in Fig. \ref{Figure_1}b. The accretion state goes into the MAD state when the saturated value of $(\dot{\phi}_{\rm acc})$ reaches a critical value. The general consensus is that the MAD state is achieved when the critical value of $\dot{\phi}_{\rm acc} \gtrsim 50$ in Gaussian units \citep{Tchekhovskoy-etal11, Narayan-etal12, Dihingia-etal21, Hong-Xuan-etal23, Aktar-etal24}. Here, we observe that the $\beta_0 =10$ model enters into MAD state at $t \gtrsim 9500 t_g$. Also, the model $\beta_0=25$ goes into the MAD state at $t \gtrsim 21000 t_g$. On the other hand, the model $\beta_0 = 50$ remains in the high magnetized SANE state, and the model $\beta_0 = 100$ represents the low magnetized SANE state $(\dot{\phi}_{\rm acc} < 50)$. We also investigate the variability of mass accretion rate and normalized magnetic flux following \citet{Narayan-etal-22}. The mean ($\mu$) and its variance around the mean ($\sigma^2$) can be defined as $\mu = \frac{1}{N} \sum_{i=1}^{N} x_i$ and $\sigma^2 = \frac{1}{N-1} \sum_{i=1}^{N} (x_i - \mu)^2$. Here $N$ is number of data points. We find that the variability for $\dot{M}_{\rm acc}$ and $\dot{\phi}_{\rm acc}$ are 2.0483 and 1.1010 for MAD state ($\beta_0 = 10$), respectively. We observe that the variability is much higher for our model compared to the GRMHD simulation in \citet{Narayan-etal-22} because our simulation model has a much higher resolution.


\subsection{Comparison of MAD and SANE state}
\label{com_MAD_SANE}

Now, we investigate the behavior of the MAD state and the transition from SANE to MAD state. In Fig. \ref{Figure_2}, we present the various temporal snapshots for the MAD state of $\beta_0 =10$. The upper row and lower row in Fig. \ref{Figure_2} represent the distribution of gas density $(\rho)$ and plasma-$\beta$ ($\beta$) parameter, respectively. The velocity field lines are indicated by grey lines. Here, we consider the time evolution of torus at $t= 7500, 9500, 25000$, and 50000 $t_g$, respectively. At the early simulation time, the accretion rate tends to increase and the flow velocity streamlines remain smooth in nature towards the horizon, as depicted by velocity streamlines in the first column of Fig. \ref{Figure_2} at time $t= 7500 t_g$. The accretion flow remains in the SANE state at time $t \leq 9000 t_g$ as normalized magnetic flux is also below the threshold value ($\dot{\phi}_{\rm acc} < 50$) of MAD state, shown in Fig. \ref{Figure_1}b for $\beta_0 = 10$. Also, the magnetic pressure ($\beta$) remains very low at the horizon at time $t = 7500 t_g$, as depicted in the lower row, the first column of Fig. \ref{Figure_2}. We also represent the transition time, when the magnetic flux crosses the critical value to become a MAD state, shown in second column at time $t= 9500 t_g$. It is found that the velocity field streamlines start to deviate from a smooth connection to the horizon and the disk region as magnetic pressure increases near the horizon. With time, enough magnetic flux accumulated at the horizon to resist further increase of accretion rate, and the accretion flow becomes MAD in nature, as shown by plasma-$\beta$ distribution in Fig. \ref{Figure_2} at times $t=25000 t_g$ and $t=50000 t_g$. The increase of magnetic pressure $(\beta << 1)$ near the horizon is also observed, as shown in Fig. \ref{Figure_2} at time $t=25000 t_g$ and $t=50000 t_g$. Moreover, it is observed that the velocity field lines of the inflowing streamlines turn outward or vertically into a prominent outflow component in the MAD state \citep{Aktar-etal24}. 

Now, we compare the 2D distribution of the other flow variables for MAD and SANE state in Fig. \ref{Figure_3}. The upper and lower rows represent the MAD ($\beta_0 = 10$) and SANE state ($\beta_0 = 100$), respectively. Here, we compare the distribution of temperature $(T)$, magnetization parameter $(\sigma_{\rm M})$, radiation energy density $(E_{\rm rad})$ and magnitude of vertical velocity ($|v_z|$) in the first, second, third and fourth columns for MAD and SANE state, respectively at time $t=50000 t_g$. The initial dense torus with a temperature of approximately $\sim 10^9$ K is surrounded by hot rarefied gas with a temperature of around $\sim 10^{13}$ K, as illustrated in Figure \ref{Figure_01}b. As time progresses, MRI activity causes an increase in magnetic pressure and synchrotron emission above and below the equator near the horizon. As time further increases, the magnetic field is significantly amplified by MRI in the MAD state compared to the SANE state. As the MAD state approaches its final stage, the torus expands increasingly and a high-velocity jet forms in the funnel region. Consequently, no remnant hot gas is left in the MAD state. However, for the SANE state, the magnetic pressure remains relatively low due to the initially weak magnetic field. The overall flow is smooth, and the initial hot rarefied gas still exists outside the torus for SANE state. We observe that the magnetization parameter is very high ($\sigma_{\rm M} \gtrsim 10$) near the funnel region for the MAD state. Conversely, we find that the magnetization parameter is much lower ($\sigma_{\rm M} <<1$) for SANE. The radiation energy density is much higher ($\gtrsim 10^3$) in the MAD state compared to the SANE state. In general, a high magnetization parameter ($\sigma_{\rm M} \gtrsim 1$) indicates the ejection of relativistic jets \citep{Dihingia-etal21, Dihingia-etal22, Narayan-etal-22, Hong-Xuan-etal23}. To investigate further, we present the distribution of vertical velocity magnitude $(|v_z|)$ in the fourth column in Fig. \ref{Figure_3}. Interestingly, we find that the vertical velocity is very high $(|v_z| \lesssim c)$ in the funnel region and extends towards the vertical outer boundary for the MAD state. However, the vertical velocity remains at low values (${|v_z|}\lesssim 0.01 c$) for the SANE state. This indicates that the MAD state is capable of producing highly relativistic magnetized jets or mass outflows \citep{Tchekhovskoy-etal11, Narayan-etal12, Dihingia-etal21, Hong-Xuan-etal23, Aktar-etal24}. On the other hand, the SANE state fails to produce high relativistic jets and only produces non-relativistic magnetized mass outflows, as shown in Fig. \ref{Figure_3} \citep{Aktar-etal24}.

\begin{figure*}
	\begin{center}
        \includegraphics[width=0.25\textwidth]{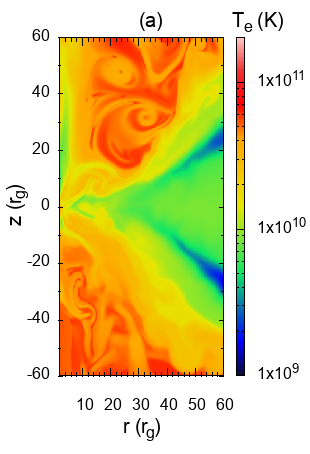} 
        \hskip -0.10 cm
        \includegraphics[width=0.25\textwidth]{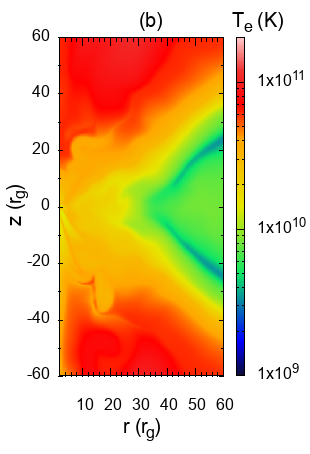} 
        \hskip -0.10 cm
        \includegraphics[width=0.25\textwidth]{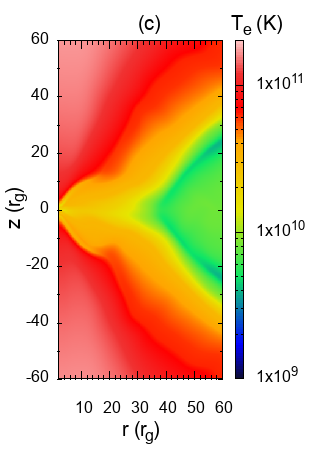} 
        \hskip -0.10 cm
	\includegraphics[width=0.25\textwidth]{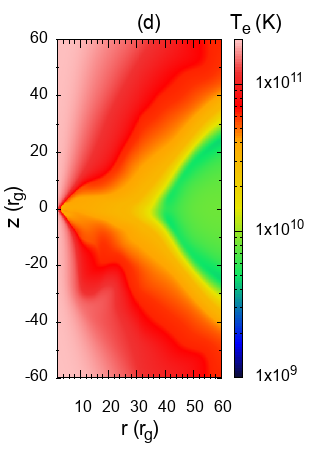} 
	\end{center}
\caption{Comparison of electron temperature ($T_e$) for $(a)$: $\beta_0 = 10$, $(b)$: $\beta_0 =25$, $(c)$: $\beta_0 =50$ and $(d)$: $\beta_0 =100$ at the simulation time $t=50000 t_g$, respectively. See the text for details.}
	\label{Figure_4}
\end{figure*}

\begin{figure*}
	\begin{center}
        \includegraphics[width=0.25\textwidth]{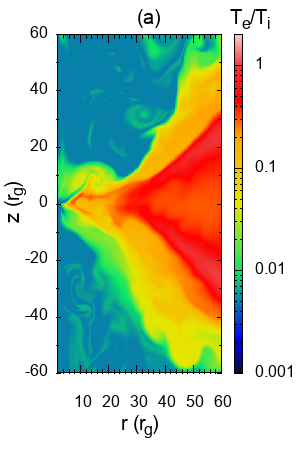} 
        \hskip -0.10 cm
        \includegraphics[width=0.25\textwidth]{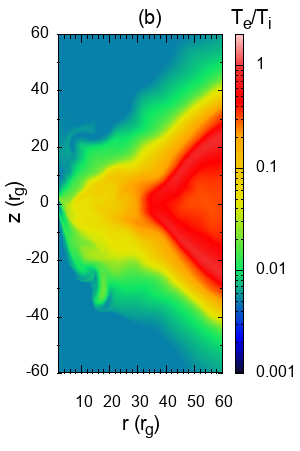} 
        \hskip -0.10 cm
        \includegraphics[width=0.25\textwidth]{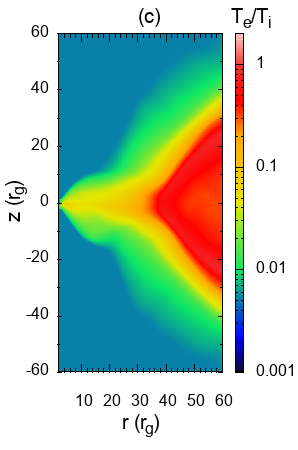} 
        \hskip -0.10 cm
	\includegraphics[width=0.25\textwidth]{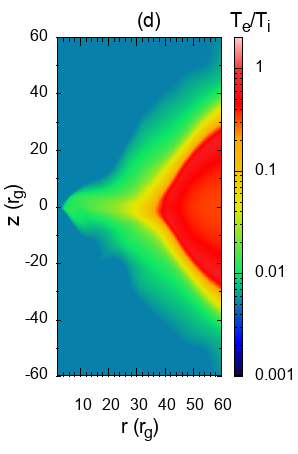}
	\end{center}
\caption{Comparison of the ratio of electron temperature to the ion temperature ($T_e/T_i$) for $(a)$: $\beta_0 = 10$, $(b)$: $\beta_0 =25$, $(c)$: $\beta_0 =50$ and $(d)$: $\beta_0 =100$ at the simulation time $t=50000 t_g$, respectively. See the text for details.}
	\label{Figure_5}
\end{figure*}

\subsubsection{Two-temperature model}
\label{two_temp_model}

It is to be mentioned that in our Rad-RMHD module of PLUTO, we use a single-temperature model, i.e. the electron temperature ($T_e$) is equal to the ion temperature ($T_i$). In this work, we adopt a self-consistent two-temperature model to evaluate the synchrotron and bremsstrahlung luminosity following the same prescription as \citet{Okuda-etal23}. To estimate the electron ($T_e$) and ion ($T_i$) temperature, we solve the radiation energy equilibrium equation in which Coulomb interaction transfer energy rate $(Q^{ie})$ from ion to electron is equal to the sum of synchrotron cooling rate ($Q_{\rm syn}$) and bremsstrahlung cooling rate ($Q^{ie}_{\rm br}$) in electrons. Therefore,
\begin{align}\label{tot_cooling_rate}
Q^{ie} = Q_{\rm syn} + Q_{\rm br}.
\end{align}
Here, the coulomb interaction rate $(Q^{ie})$ is given by following \citet{Stepney-Guilbert-83} as
\begin{align} \label{Syn_rate}
\begin{split}
Q^{ie} & = 5.61 \times 10^{-32} \frac{n_e n_i (T_i -T_e)}{K_2(1/\Theta_e) K_1(1/\Theta_i)} \\
       & \times \left[ \frac{2(\Theta_e+\Theta_i)^2+1}{(\Theta_e+\Theta_i)} K_1 \left(\frac{\Theta_e+\Theta_i}{\Theta_e \Theta_i}\right) + 2K_0 \left(\frac{\Theta_e+\Theta_i}{\Theta_e \Theta_i}\right) \right] \\
       & {\rm erg~cm^{-3}~s^{-1}},
\end{split}
\end{align}
where $n_e$ and $n_i$ are the number density of electrons and ions. $K_0$, $K_1$ and $K_2$ are the modified Bessel functions. Also, $\Theta_e = \frac{k_{\rm B}T_e}{m_e c^2}$ and $\Theta_i = \frac{k_{\rm B}T_i}{m_p c^2}$ are the dimensionless electron and ion temperature, respectively. Here $k_{\rm B}$, $m_e$, $m_p$ and $c$ are Boltzmann constant, electron mass, proton mass, and the speed of light.

In order to calculate bremsstrahlung cooling rate, we adopt \citet{Stepney-Guilbert-83, Esin-etal96} prescription. The free-free bremsstrahlung cooling rate of ionized plasma consists of electrons and ions and is given by
\begin{align} \label{Brem_rate}
Q_{\rm br} = Q^{ie}_{\rm br} + Q^{ee}_{\rm br},
\end{align}
where
\begin{align}
Q^{ie}_{\rm br} = 1.48 \times 10^{-22} n_e^2 F^{ie} (\Theta_e)~~~~ {\rm erg~ cm^{-3}~ s^{-1}},
\end{align}
where
\[
    F^{ie}(\Theta_e)= 
\begin{cases}
    4 \sqrt{\frac{2 \Theta_e}{\pi^3}} (1+1.781\Theta_e^{1.34}), & \text{if } \Theta_e < 1\\
    \frac{9 \Theta_e}{2 \pi}[\ln({1.12 \Theta_e + 0.48})+1.5],  & \text{if } \Theta_e > 1.
\end{cases}
\]
Also, we have
\begin{align}
\begin{split}
Q^{ee}_{\rm br} & =
\begin{cases}
    2.56 \times 10^{22} n_e^2 \Theta_e^{1.5}(1+1.10\Theta_e+\Theta_e^2 -1.25\Theta_e^{2.5}), & \text{if } \Theta_e < 1\\
    3.40 \times 10^{-22} n_e^2 \Theta_e [\ln(1.123 \Theta_e)+1.28],  & \text{if } \Theta_e > 1.\\
\end{cases}\\
& {\rm erg~cm^{-3}~s^{-1}}.
\end{split}
\end{align}
Further, in the presence of a magnetic field, the hot electrons radiate via a thermal synchrotron process. The synchrotron cooling rate is given by following \citet{Narayan-Yi95, Esin-etal96} as
\begin{align}
  \begin{aligned}
    Q_{\rm syn} = & \frac{2\pi k_{\rm B} T_i \nu_c^3}{3Hc^2} + 6.76\times10^{-28}\frac{n_e}{K_2\bigg(\frac{1}{\Theta_e}\bigg)a_1^{1/6}}\\
                 & \times\bigg[\frac{1}{a_4^{11/2}}\Gamma\left(\frac{11}{2},a_4\nu_c^{1/3}\right) + \frac{a_2}{a_4^{19/4}}\Gamma\left(\frac{19}{4},a_4\nu_c^{1/3}\right) \\
                 &+ \frac{a_3}{a_4^4}\left(a_4^3 \nu_c + 3a_4^2\nu_c^{2/3} + 6a_4\nu_c^{1/3} +6\right)\exp(-a_4\nu_c^{1/3})\bigg]\ \\
                 & {\rm erg~cm^{-3}~s^{-1}} ,
\end{aligned}
\end{align}
where $H$ is the scale disk height and the coefficients $a_{1-4}$ are given by
\begin{align}
a_1 = \frac{2}{3\nu_0 \Theta_e^2}, ~~a_2 = \frac{0.4}{a_1^{1/4}},~~ a_3 = \frac{0.5316}{a_1^{1/2}},~~ a_4 = 1.8899a_1^{1/3}
\end{align}
where, $\nu_0 = \frac{eB}{2 \pi m_e c}$ and $\nu_c = \frac{3}{2}\nu_0 \Theta_e^2 x_{\rm M}$ are the characteristic synchrotron frequencies. Here, $e$ and $B$ are the electron charge and strength of the magnetic field. $x_{\rm M}$ can be obtained from the following equation
\begin{align}
  \begin{aligned}
\exp{1.8899 x_{\rm M}^{1/3}} = & 2.49 \times 10^{-10} \frac{4 \pi n_e r}{B} \frac{1}{\Theta_e^3 K_2 \bigg(\frac{1}{\Theta_e}\bigg)} \\
                               & \times \bigg(\frac{1}{x_{\rm M}^{7/6}} + \frac{0.40}{x_{\rm M}^{17/12}}+ \frac{0.5316}{x_{\rm M}^{5/3}}\bigg).
  \end{aligned}
\end{align}
Here, the ‘Gamma function’ is defined as $\Gamma(a,x)=\int_x^\infty t^{a-1}e^{-t}~dt$. 

Now, to incorporate the two-temperature model from single-temperature simulation results, we have
\begin{align}\label{tot_temp}
T_e + T_i = 2 T,
\end{align}
using $P_{\rm gas} = n k_{\rm B} T = n_i k_{\rm B} T_i + n_e k_{\rm B} T_e$ with $n = n_e + n_i$ and $n_e = n_i$, where $n$ and $T$ are the number density and temperature for single-temperature model. We numerically solve equation (\ref{tot_cooling_rate}) and (\ref{tot_temp}) to evaluate electron temperature $(T_e)$ and ion temperature $(T_i)$ for our model.

 In Fig. \ref{Figure_4}, we compare the distribution of the electron temperature ($T_e$). We present the electron temperature ($T_e$) distribution for different initial magnetic field values of $\beta_0 = 10$, 25, 50, and 100 in Fig. \ref{Figure_4}a, \ref{Figure_4}b, \ref{Figure_4}c and \ref{Figure_4}d, respectively, at the simulation time $t = 50000 t_g$. We observe that, with the increase of magnetic field (from Fig. \ref{Figure_4}d to Fig. \ref{Figure_4}a), the central density torus expands vertically towards the horizon. In the SANE state ($\beta_0 = 100$), the dense central torus with electron temperature $T_e \gtrsim 10^{9}$K is surrounded by very hot rarefied gas with electron temperature $T_e \gtrsim 10^{11}$K, as shown in Fig. \ref{Figure_4}d. In the case of the MAD state ($\beta_0 = 10$), a dense and turbulent region forms vertically near the horizon due to the high magnetic field. Surrounding the torus and disk is hot and less rarefied gas with an electron temperature of $T_e \gtrsim 10^{10}$K, as shown in Fig. \ref{Figure_4}a. We observe that the very hot and rarefied gas surrounding the funnel region persists in the SANE state. This is because the initial hot and rarefied gas is not sufficiently heated by bremsstrahlung and synchrotron radiation, and is also not disrupted by MRI turbulence, as depicted in Fig. \ref{Figure_4}d. The torus of the SANE model evolves more slowly, and it takes a long time for MRI to prevail far from the central torus. In contrast, in the MAD model, the MRI turbulence develops far from the equator. The funnel region is occupied with gas streamed out from the torus, and the initial rarefied hot gas is driven out as jets or outflows from the funnel region, as shown in Fig. \ref{Figure_4}a. In the section \ref{com_MAD_SANE}, we have already described similar features. We also compare the ratio of electron temperature to the ion temperature $(T_e/T_i)$ in Fig. \ref{Figure_5}. The temperature ratio generally depicts the distribution of thermal energy of fluid over electrons and ions. Here, we present the distribution of the temperature ratio ($T_e/T_i$) for different initial magnetic fields, $\beta_0 =10$, 25, 50, and 100 in Fig. \ref{Figure_5}a, \ref{Figure_5}b, \ref{Figure_5}c, and \ref{Figure_5}d, respectively at the simulation time $t=50000 t_g$. The initial temperature ratio $(T_e/T_i)$ ranges from $0.2-0.8$ in the torus region to $0.001-0.08$ in the outer region. The variation of $T_e/T_i$ with the increase of magnetic field is similar to the variation of $T_e$, as shown in Fig. \ref{Figure_5}a, \ref{Figure_5}b, \ref{Figure_5}c, \ref{Figure_5}d. These variations of $T_e$ and $T_e/T_i$ with the increase of magnetic field are similar to the variations of density and temperature for MAD and SANE, as shown in Fig. \ref{Figure_2} and Fig. \ref{Figure_3}.

\begin{figure}
	\begin{center}
        \includegraphics[width=0.50\textwidth]{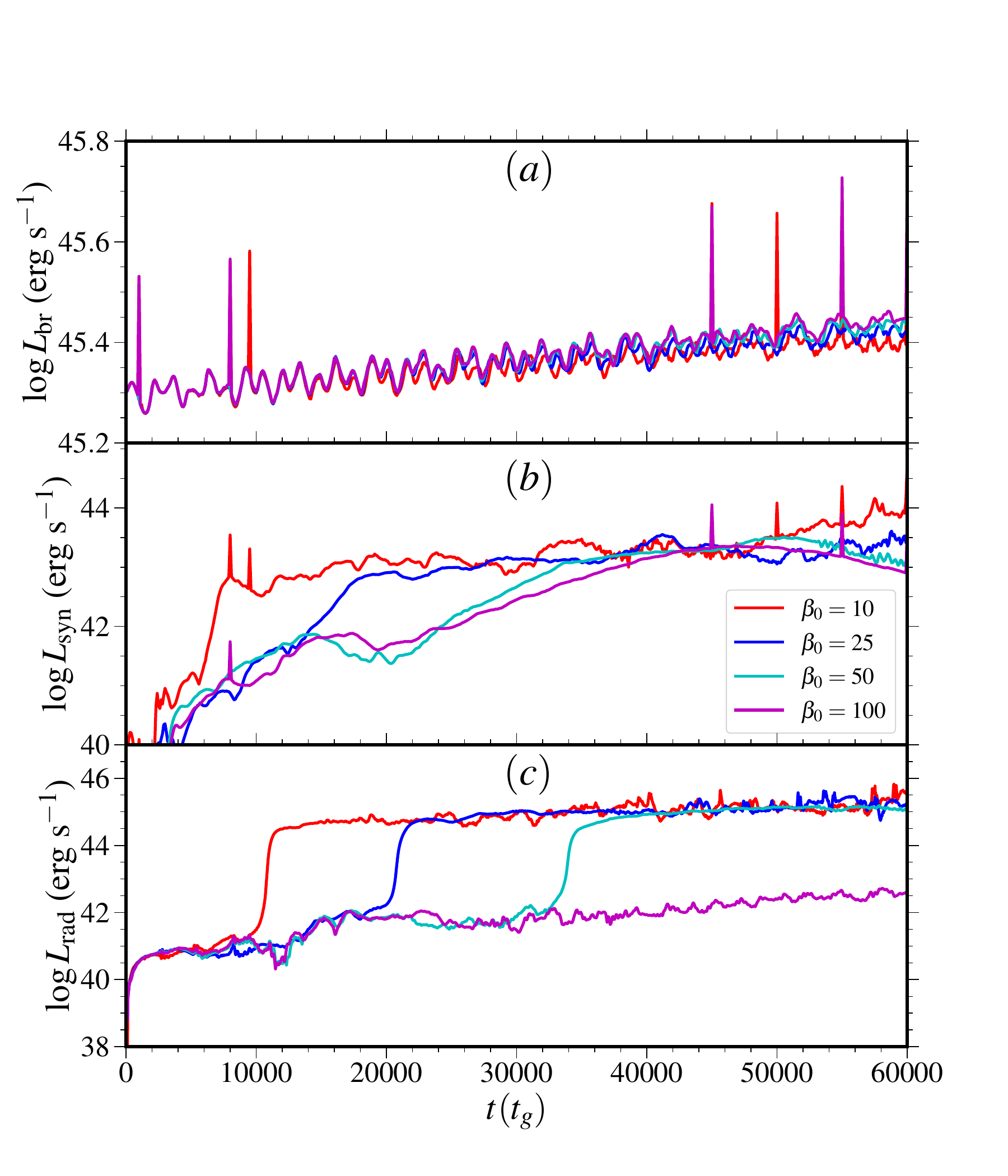} 
	\end{center}
	\caption{Variation of bremsstrahlung, synchrotron and radiative luminosity with time for different magnetic field $\beta_0 = 10, 25, 50$ and 100. See the text for details.}
	\label{Figure_6}
\end{figure}

\begin{figure}
	\begin{center}
        \includegraphics[width=0.49\textwidth]{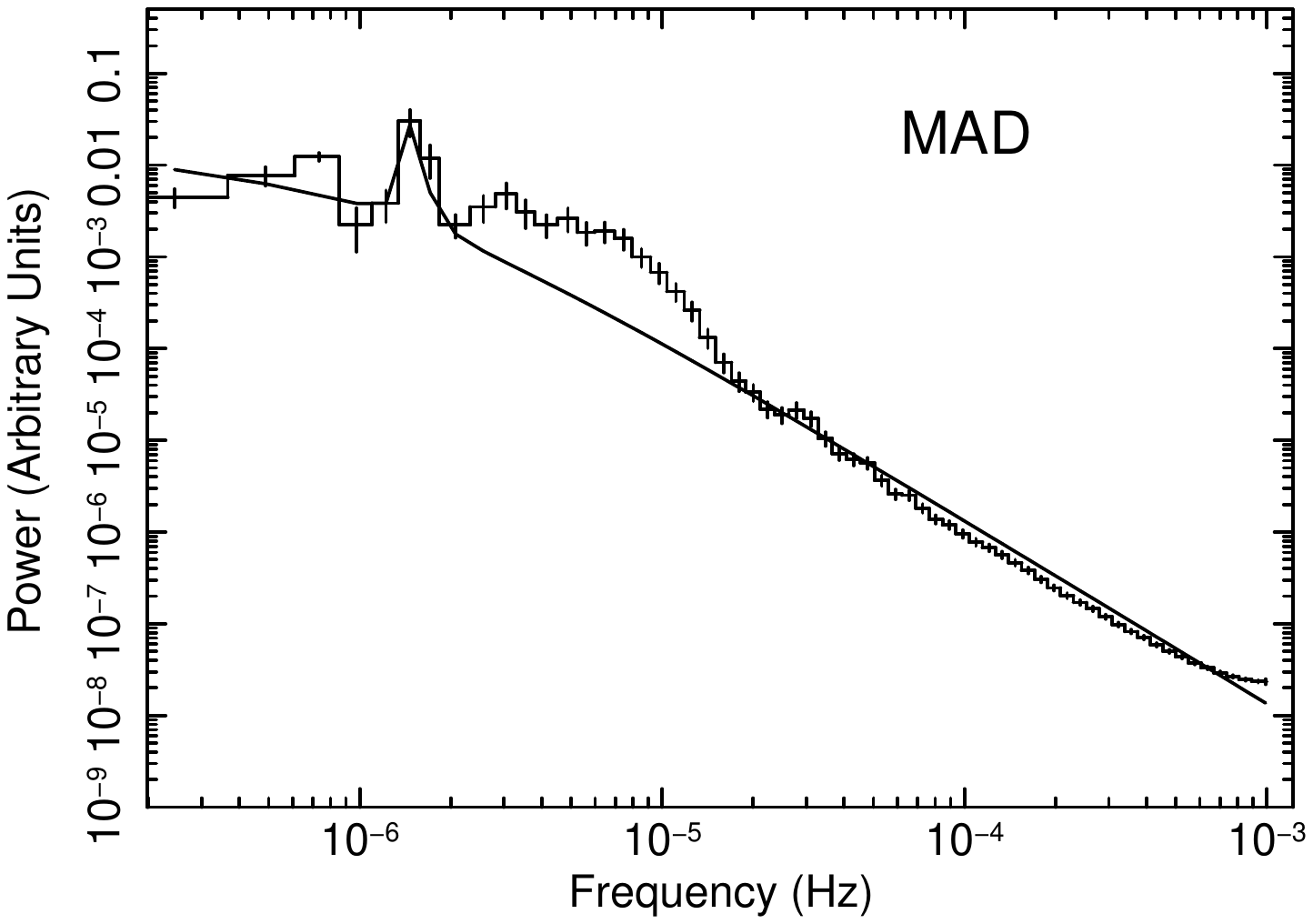} 
        \includegraphics[width=0.49\textwidth]{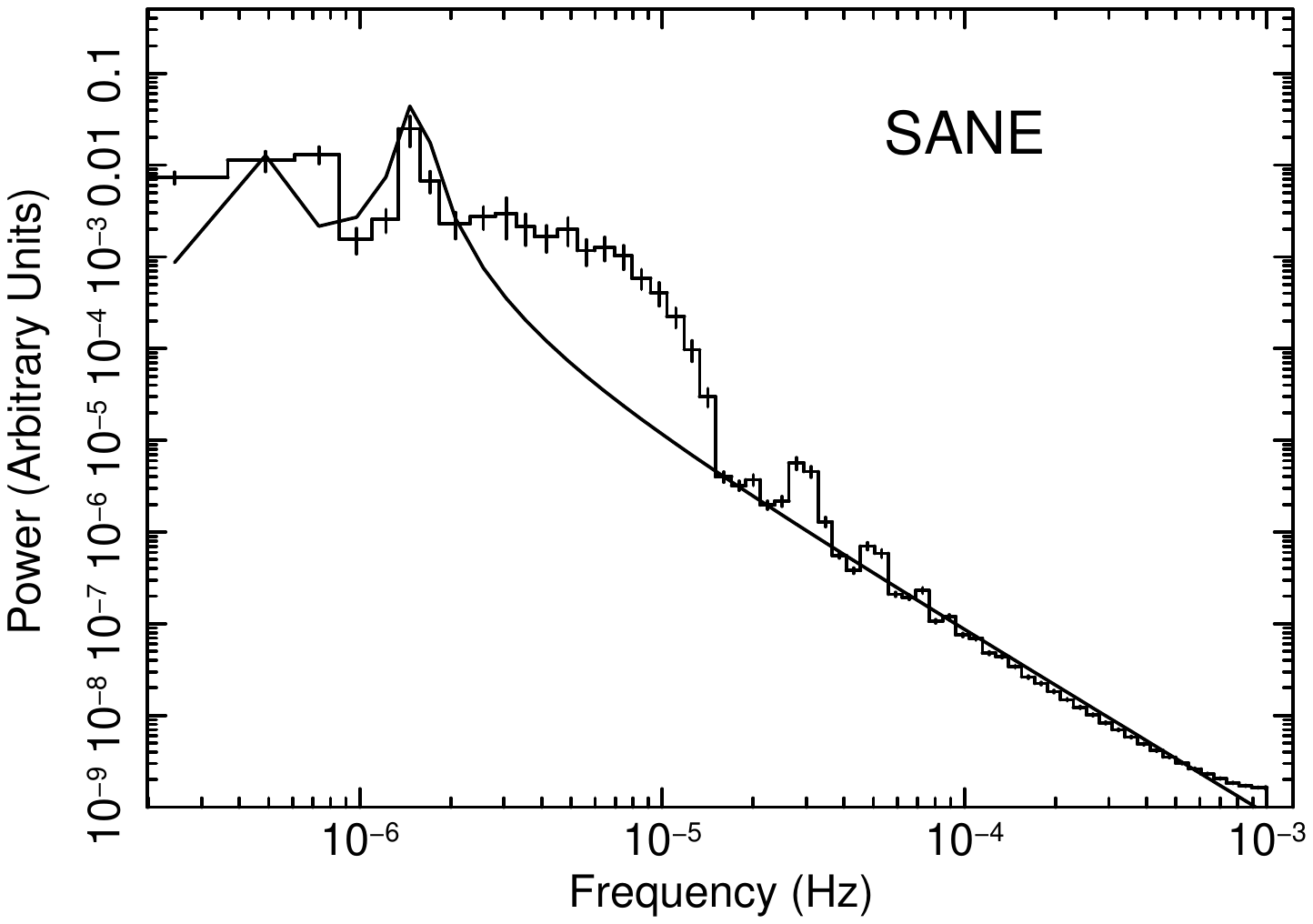} 
	\end{center}
	\caption{Power density spectrum (PDS) of bremsstrahlung luminosity variation ($L_{\rm br}$) for MAD ($\beta_0 = 10$) and SANE ($\beta_0 = 100$) state.}
	\label{Figure_7}
\end{figure}

\begin{figure}
	\begin{center}
        \includegraphics[width=0.50\textwidth]{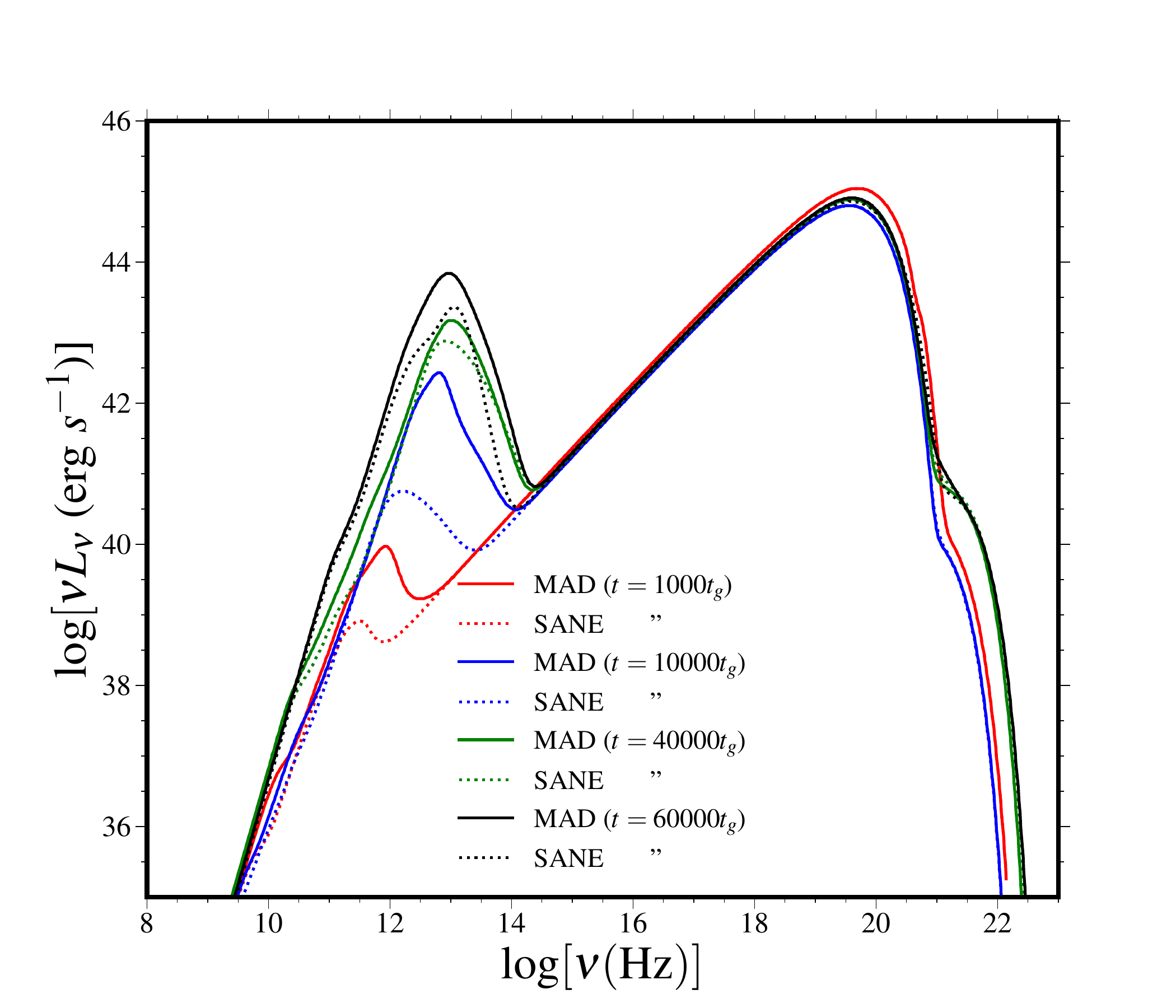} 
	\end{center}
	\caption{Evolution of spectral energy distribution (SED) of MAD ($\beta_0 = 10$) and SANE ($\beta_0 = 100$) state with simulation time. See the text for details.}
	\label{Figure_8}
\end{figure}

\begin{figure}
	\begin{center}
        \includegraphics[width=0.50\textwidth]{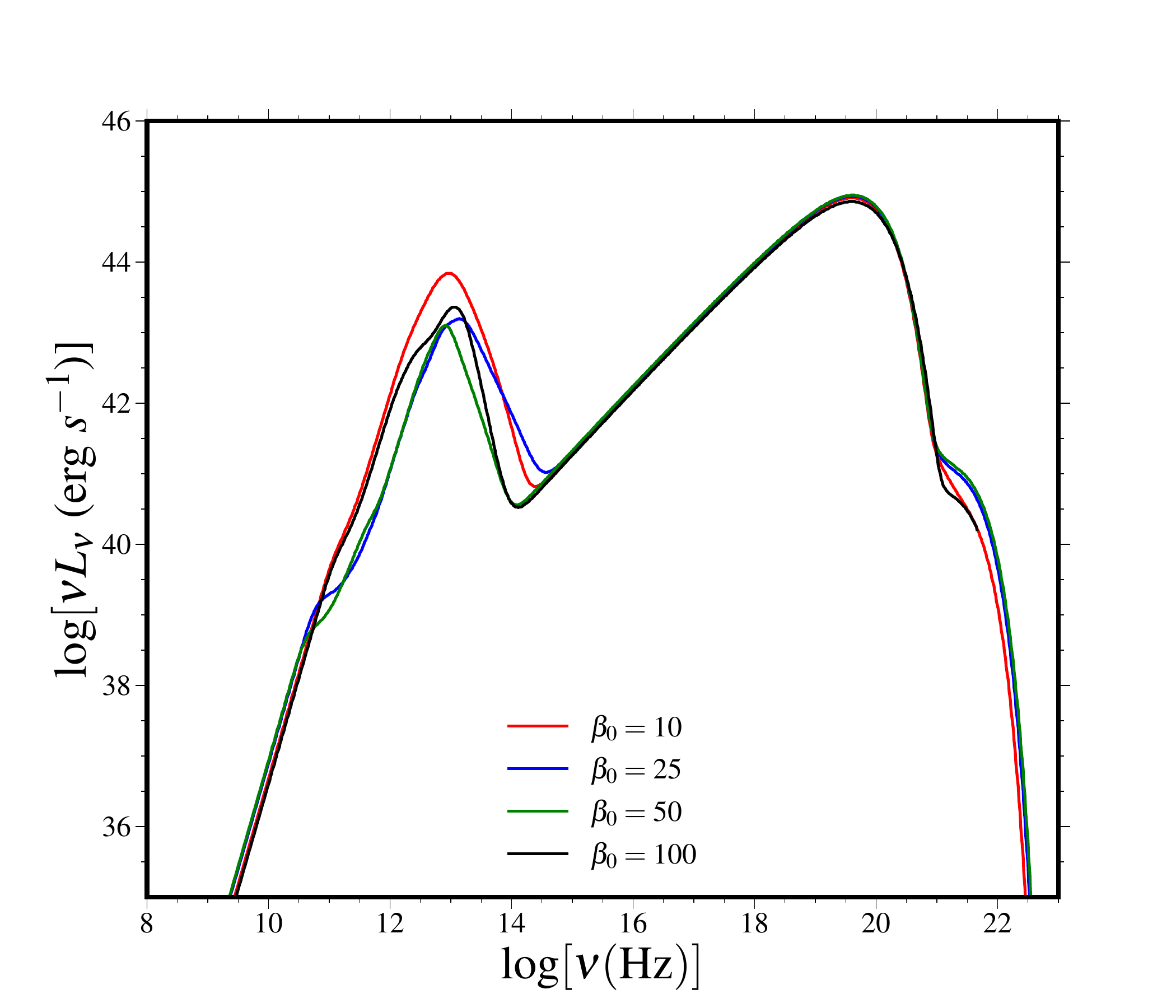} 
	\end{center}
	\caption{Comparison of spectral energy distribution (SED) by varying magnetic fields ($\beta_0$) at time $t=60000 t_g$. See the text for details.}
	\label{Figure_9}
\end{figure}

\begin{figure}
	\begin{center}
        \includegraphics[width=0.49\textwidth]{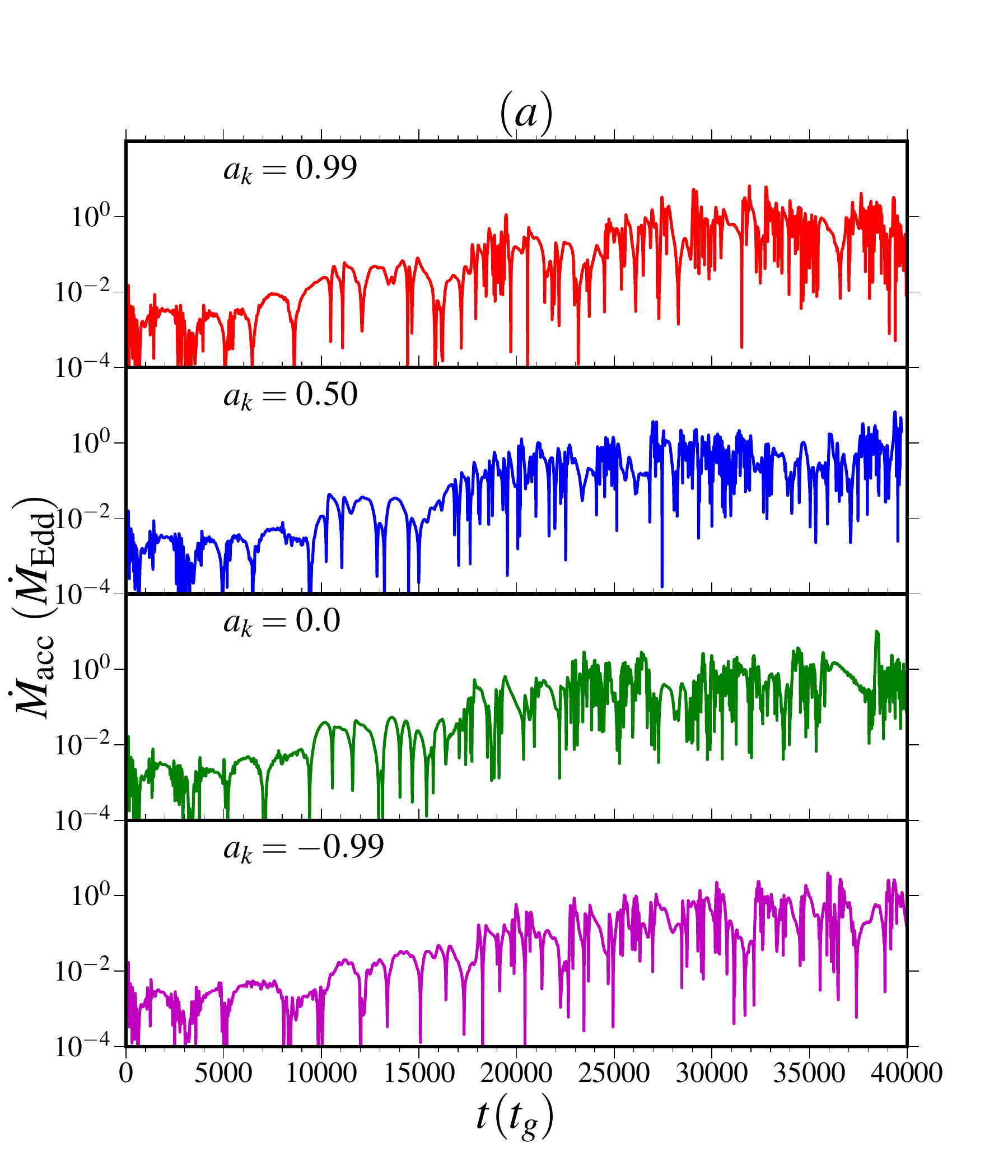} 
		\includegraphics[width=0.49\textwidth]{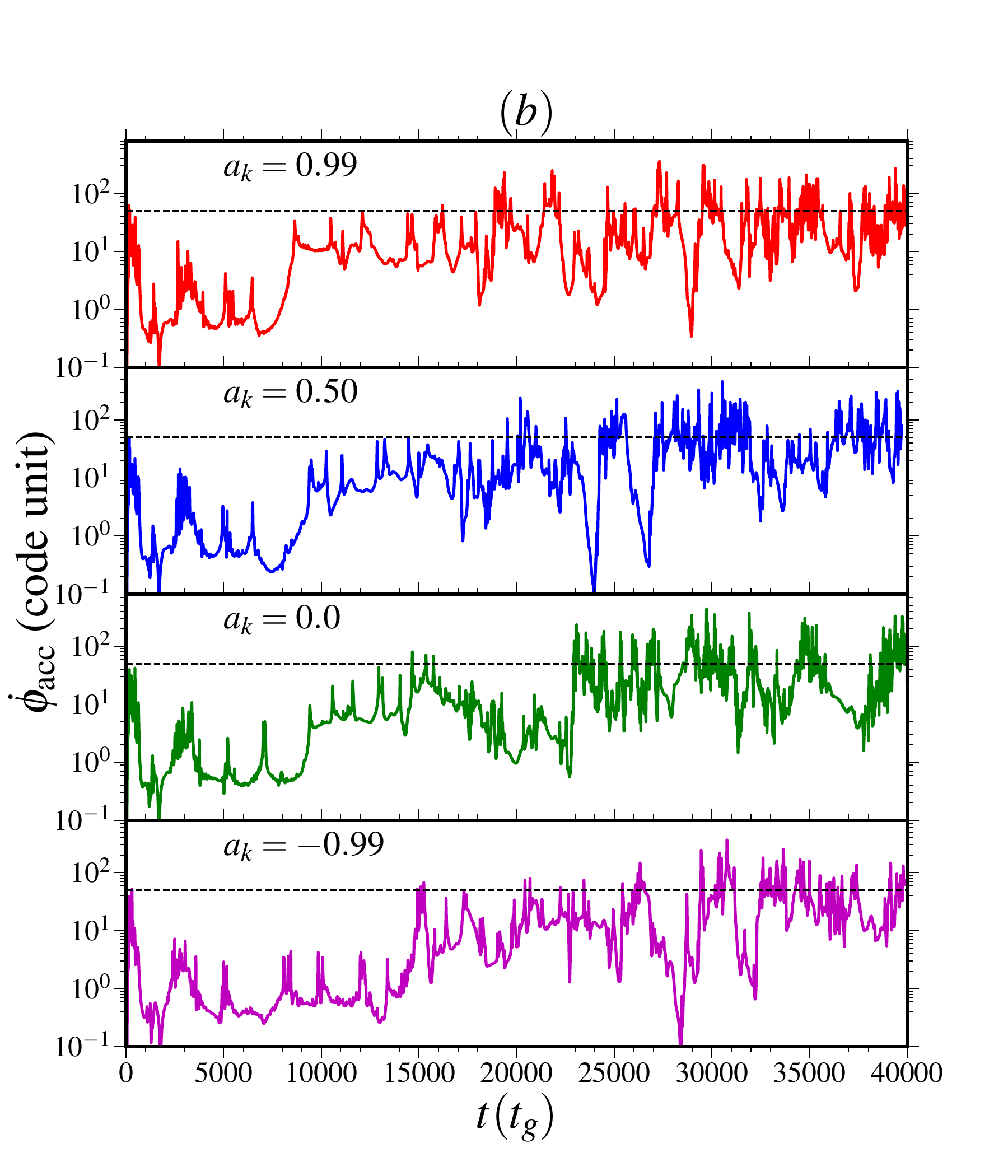} 
	\end{center}
	\caption{Temporal evolution of $(a)$: mass accretion rate $\dot{M}_{\rm acc} (\dot{M}_{\rm Edd})$
    and $(b)$: normalized magnetic ﬂux $\dot{\phi}_{\rm acc}$ in code units accumulated at the 
    black hole inner boundary with the simulation time for different black hole spin $(a_k)$. Dashed horizontal lines are for $\dot{\phi}_{\rm acc}$ = 50. Here, we consider the spin as $a_k$ = 0.99, 0.50, 0.0 and -0.99. See the text for details.} 
	\label{Figure_10}
\end{figure}

\begin{figure*}
	\begin{center}
        \includegraphics[width=0.23\textwidth]{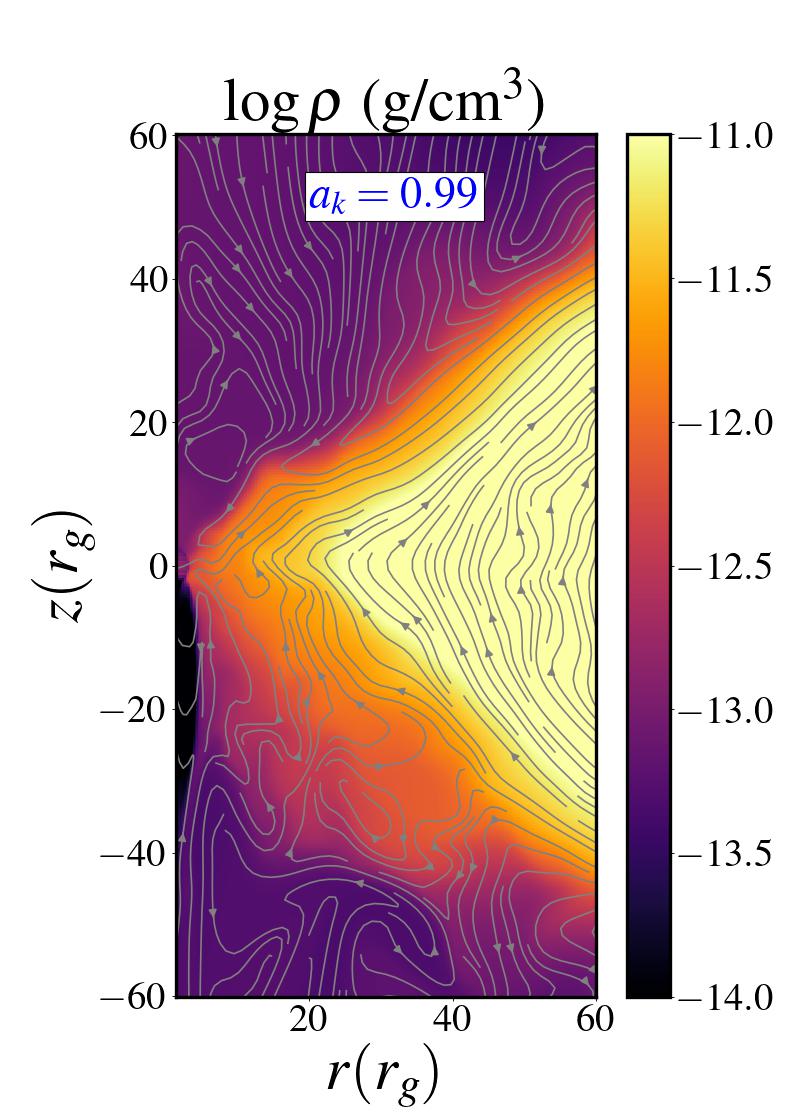} 
        \hskip -2mm
        \includegraphics[width=0.23\textwidth]{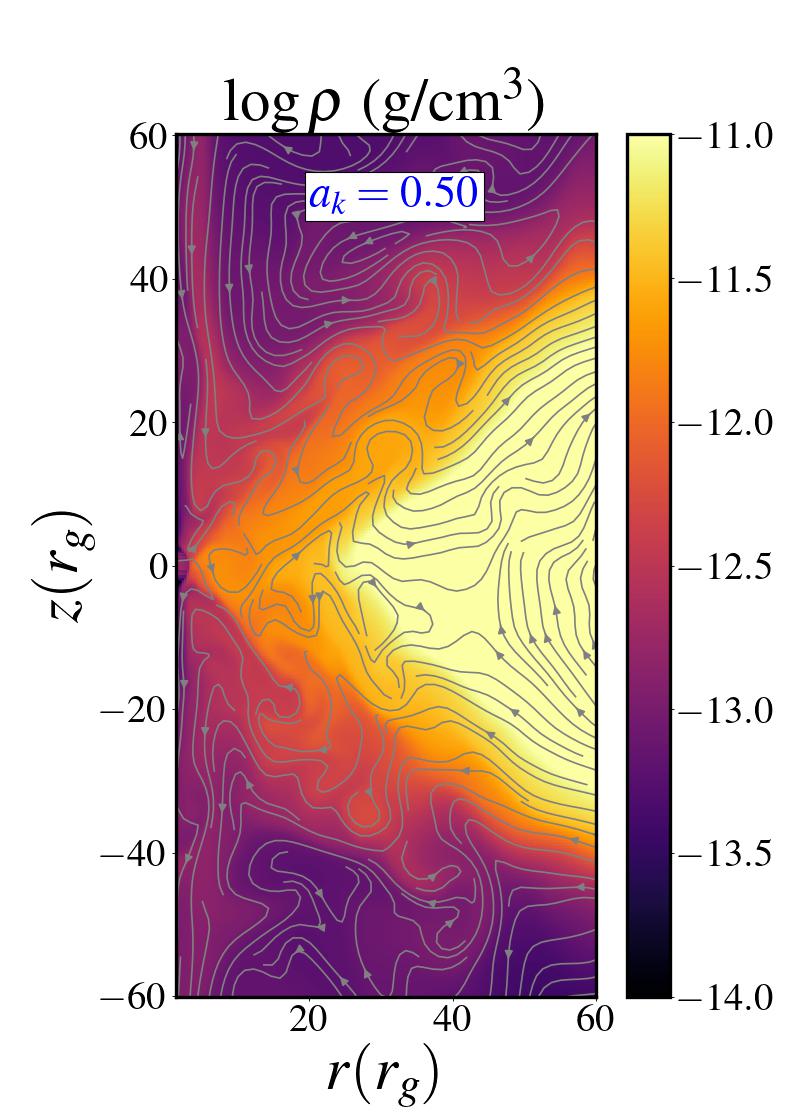} 
        \hskip -2mm
		\includegraphics[width=0.23\textwidth]{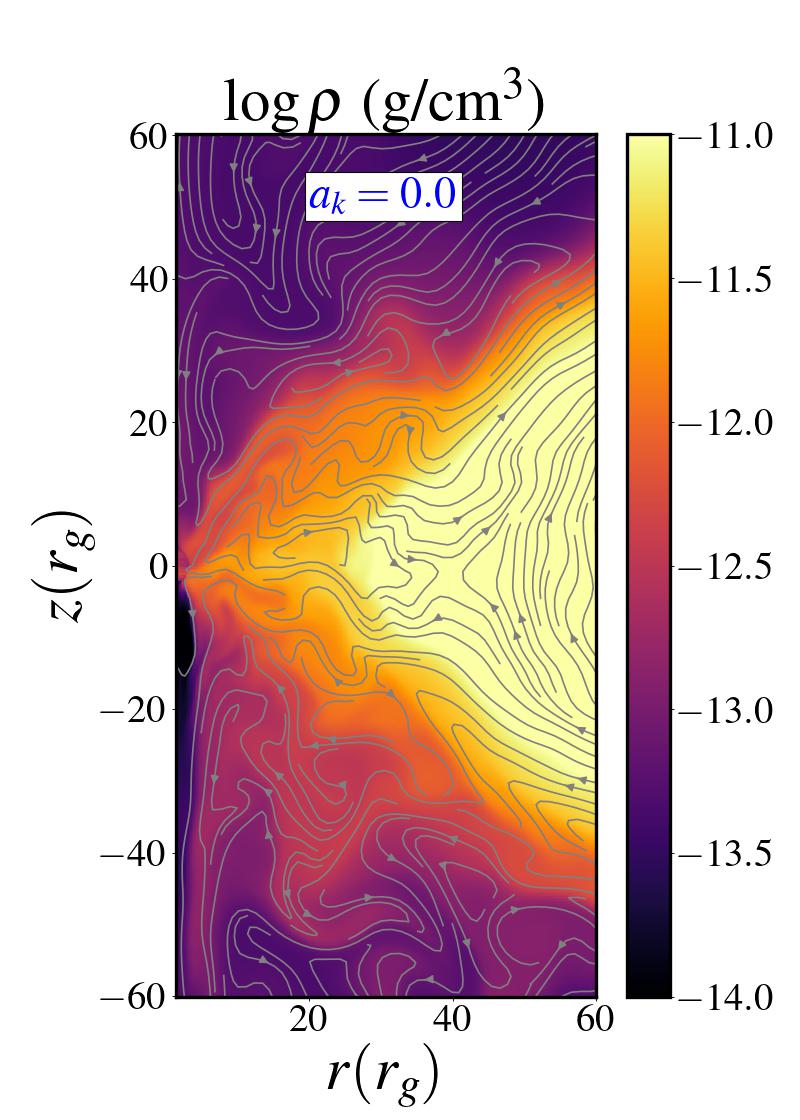} 
        \hskip -2mm
        \includegraphics[width=0.23\textwidth]{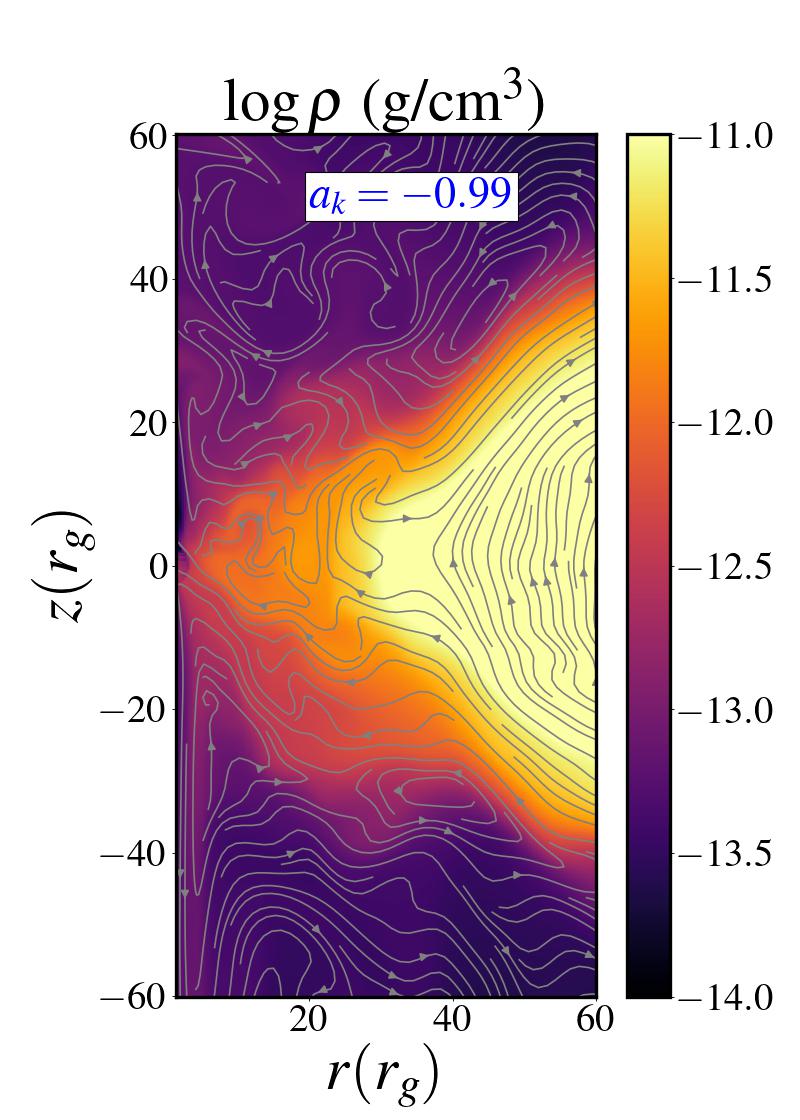} 
     
        \includegraphics[width=0.23\textwidth]{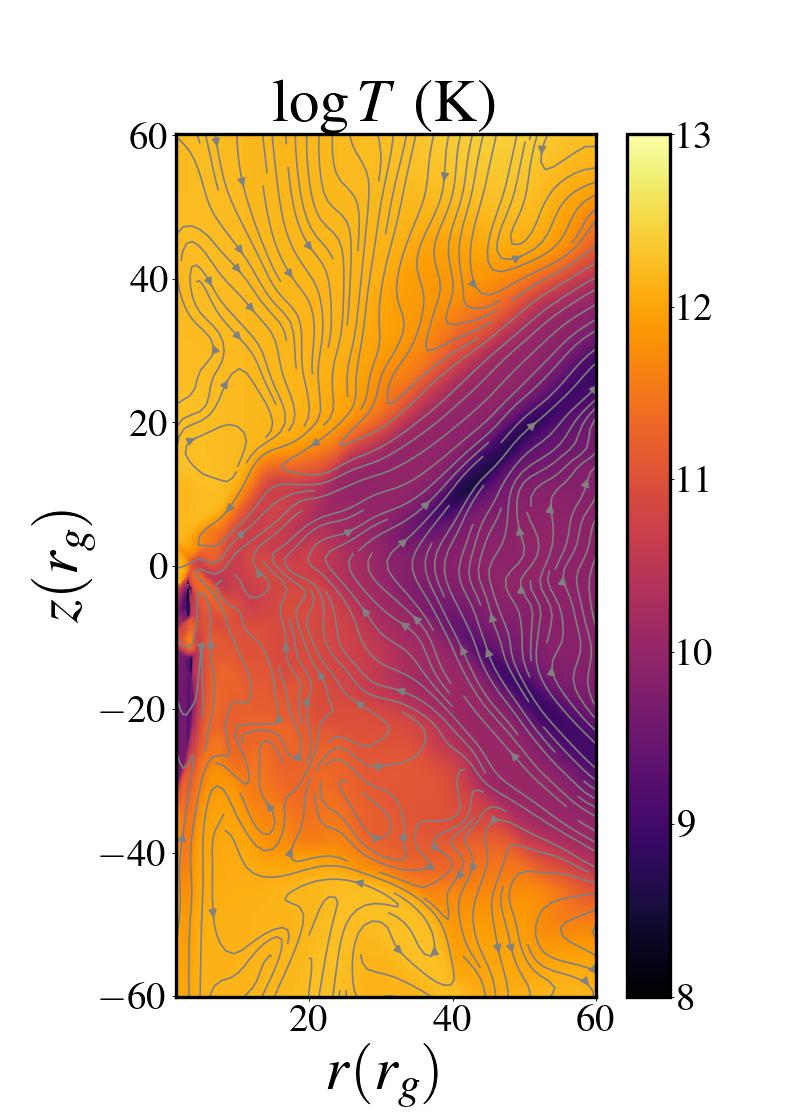} 
        \hskip -2mm
        \includegraphics[width=0.23\textwidth]{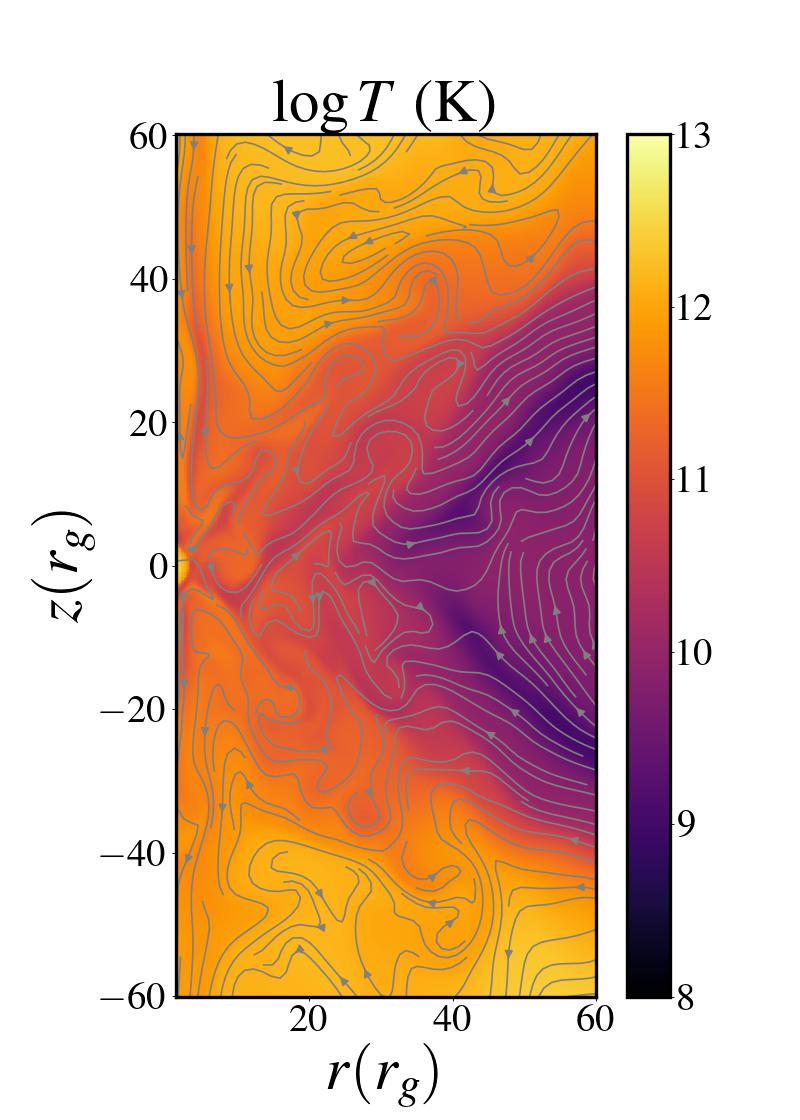} 
        \hskip -2mm
		\includegraphics[width=0.23\textwidth]{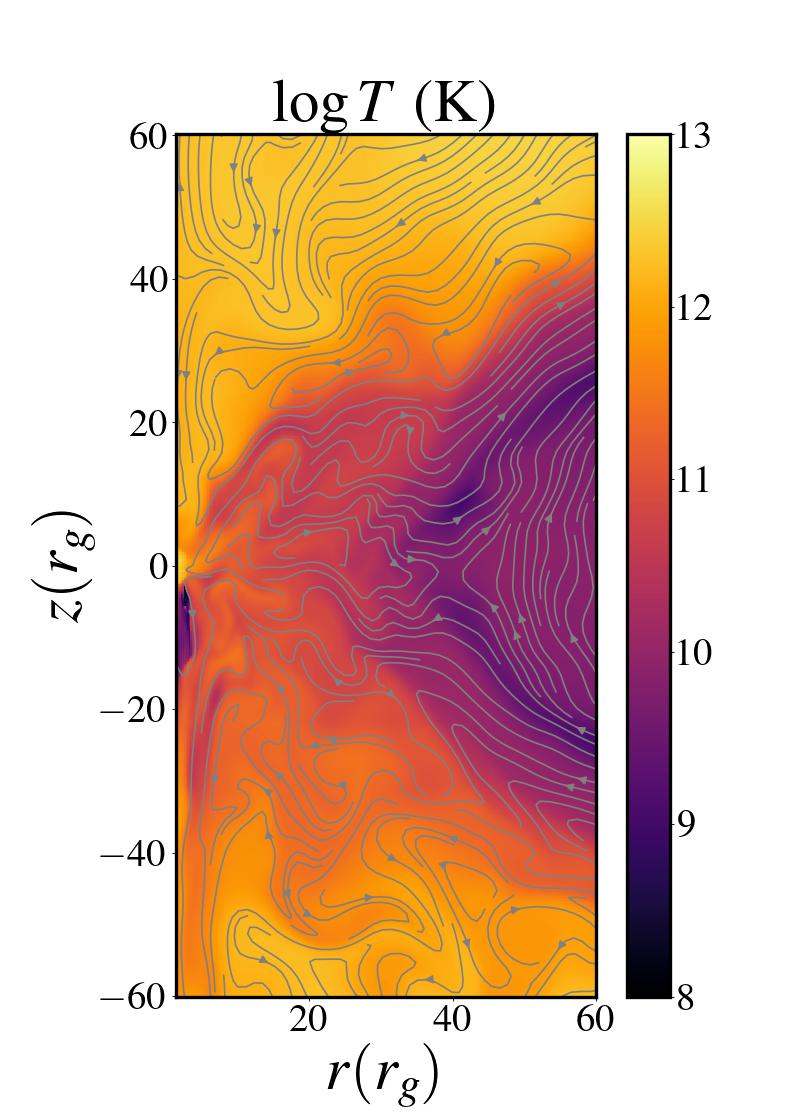} 
        \hskip -2mm
        \includegraphics[width=0.23\textwidth]{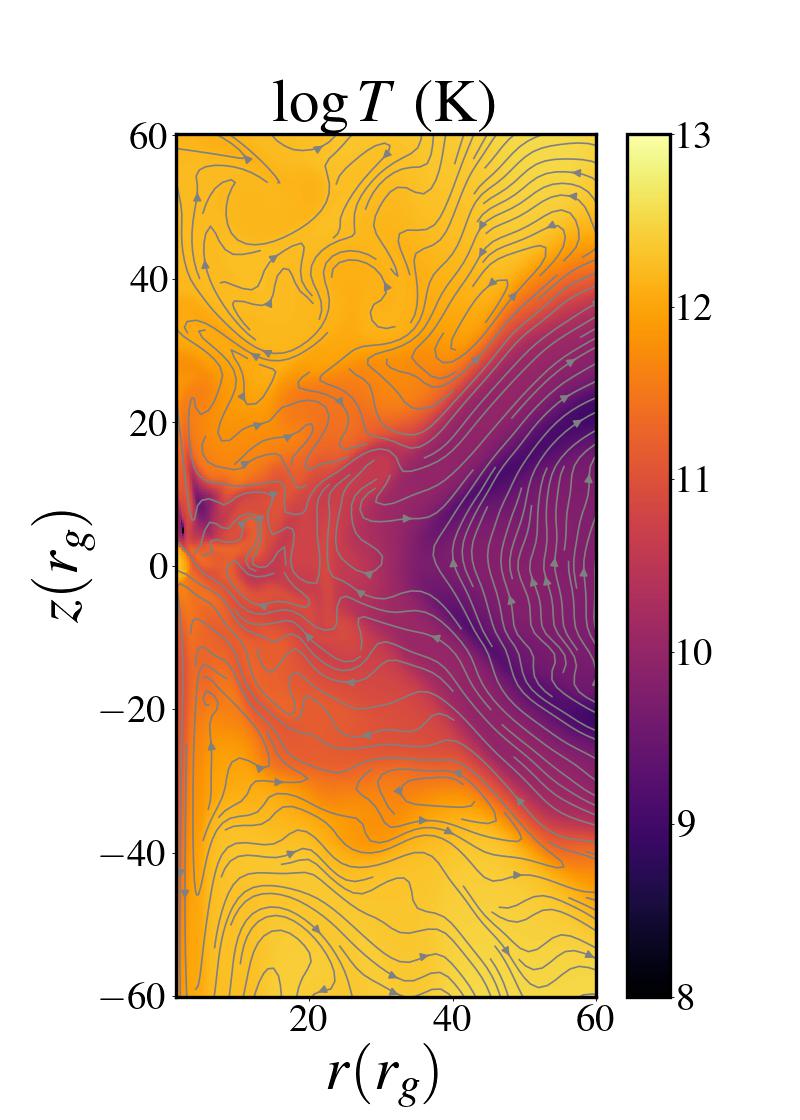} 
        
        \includegraphics[width=0.23\textwidth]{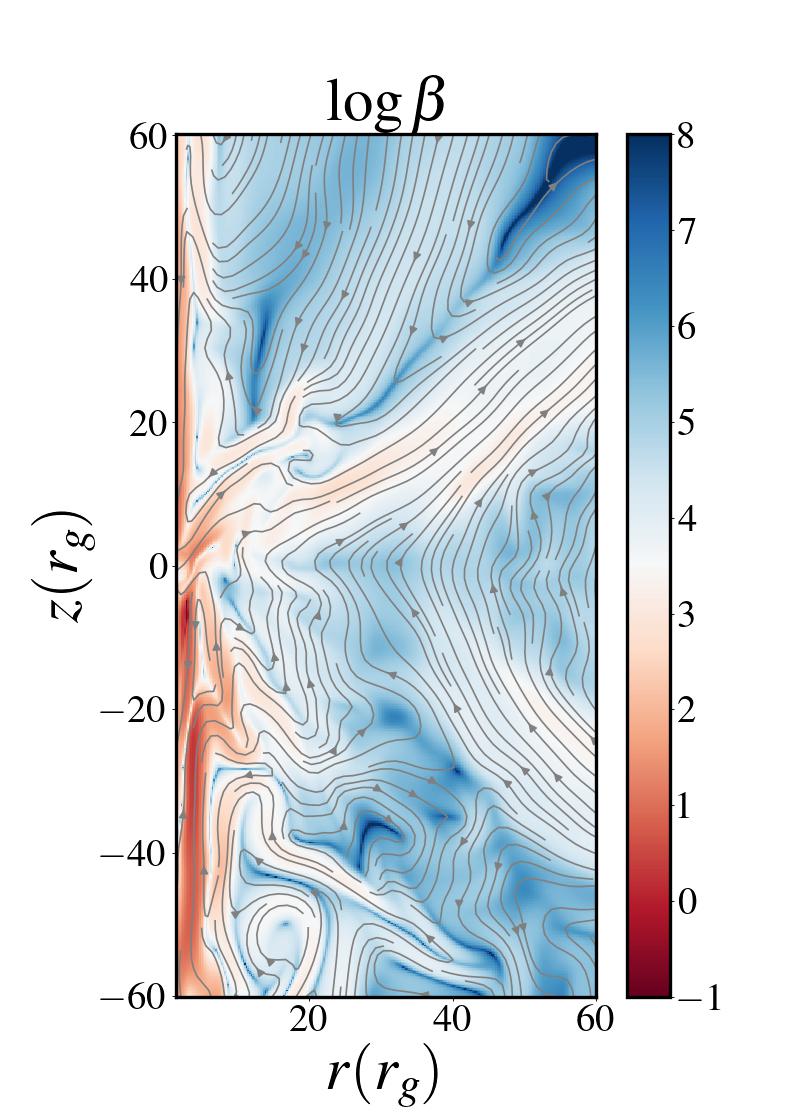} 
        \hskip -2mm
        \includegraphics[width=0.23\textwidth]{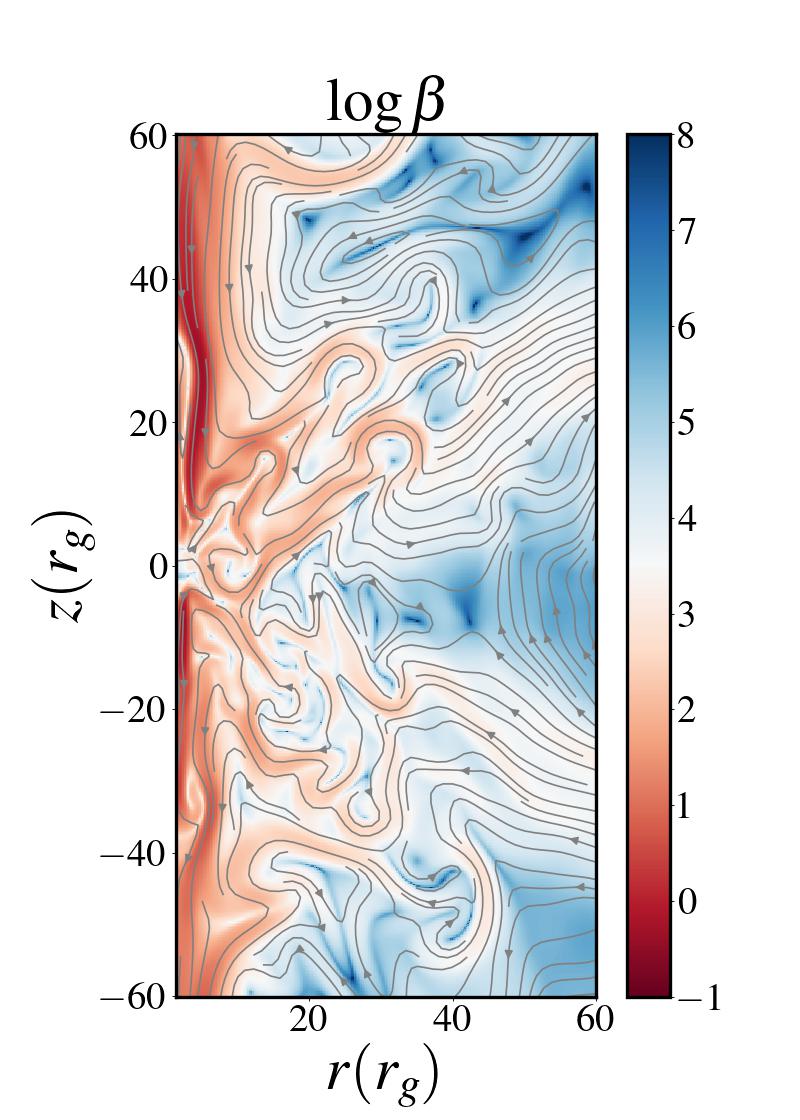} 
        \hskip -2mm
		\includegraphics[width=0.23\textwidth]{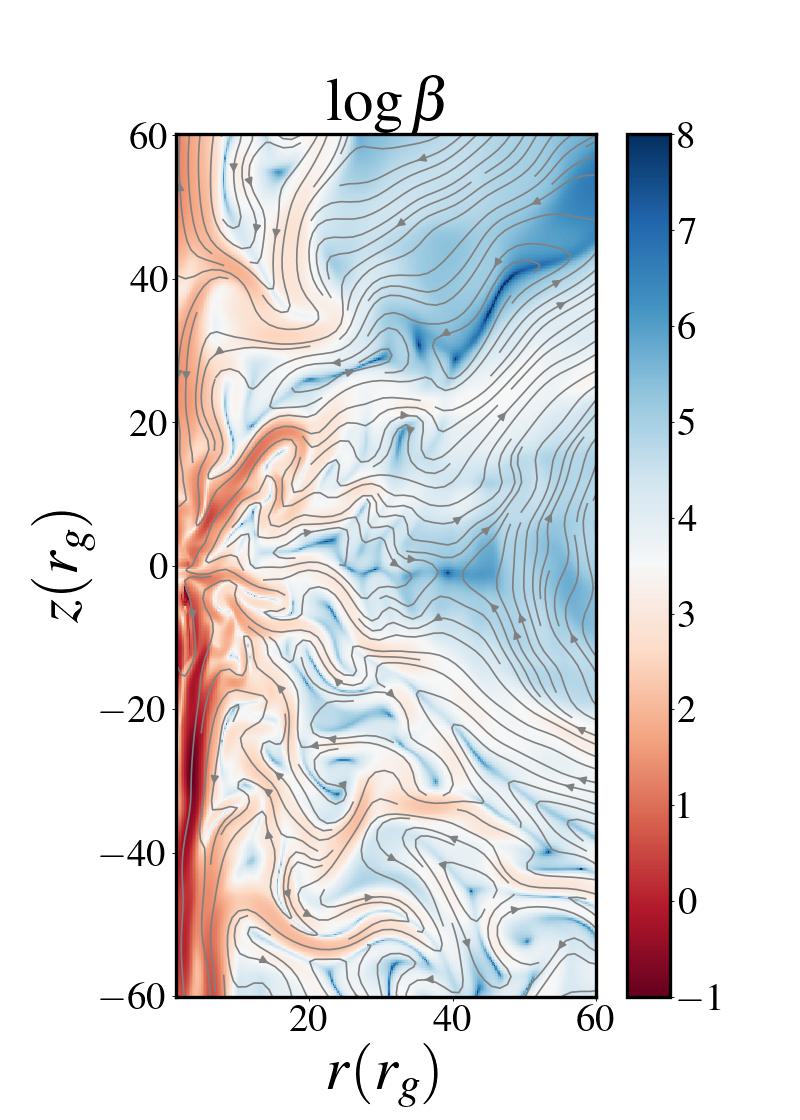} 
        \hskip -2mm
        \includegraphics[width=0.23\textwidth]{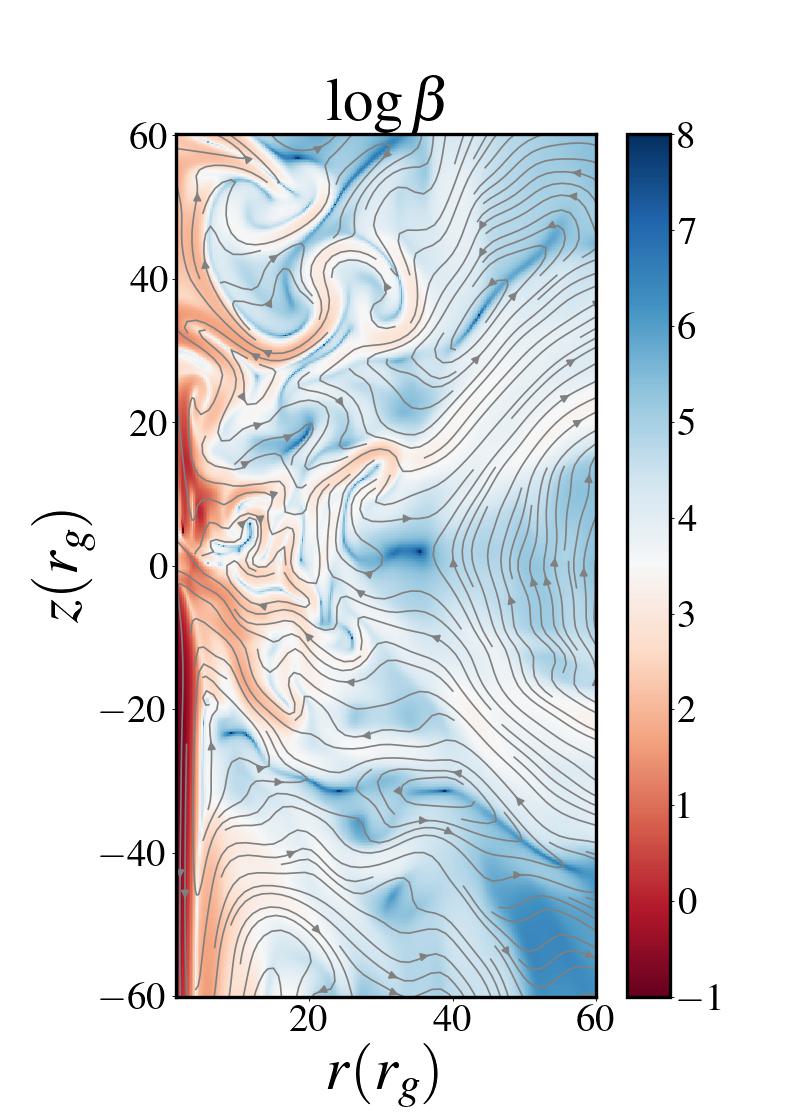} 

        \includegraphics[width=0.23\textwidth]{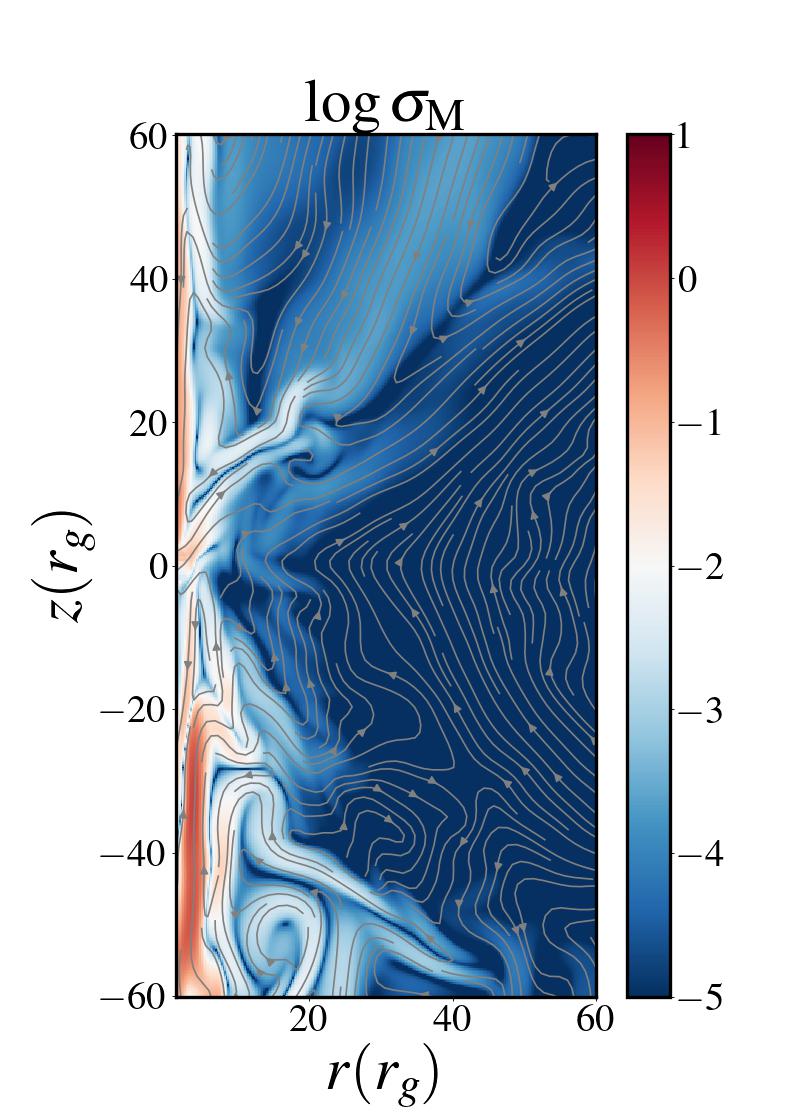} 
        \hskip -2mm
        \includegraphics[width=0.23\textwidth]{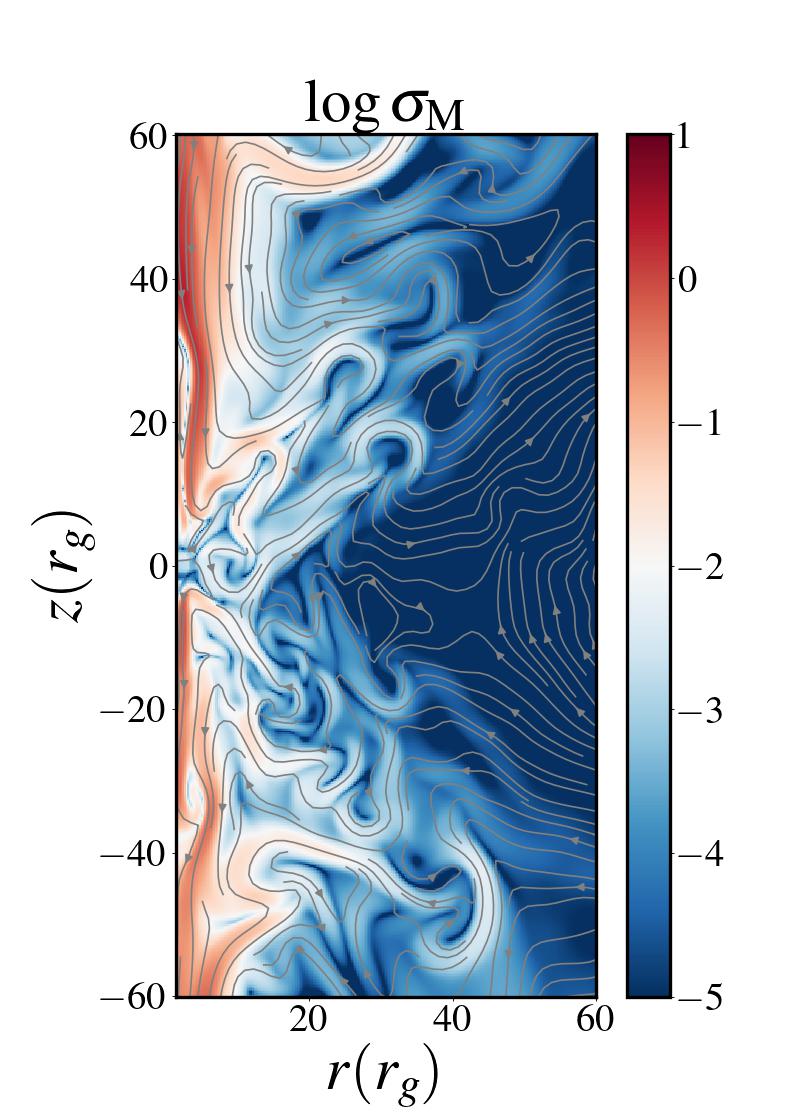} 
        \hskip -2mm
		\includegraphics[width=0.23\textwidth]{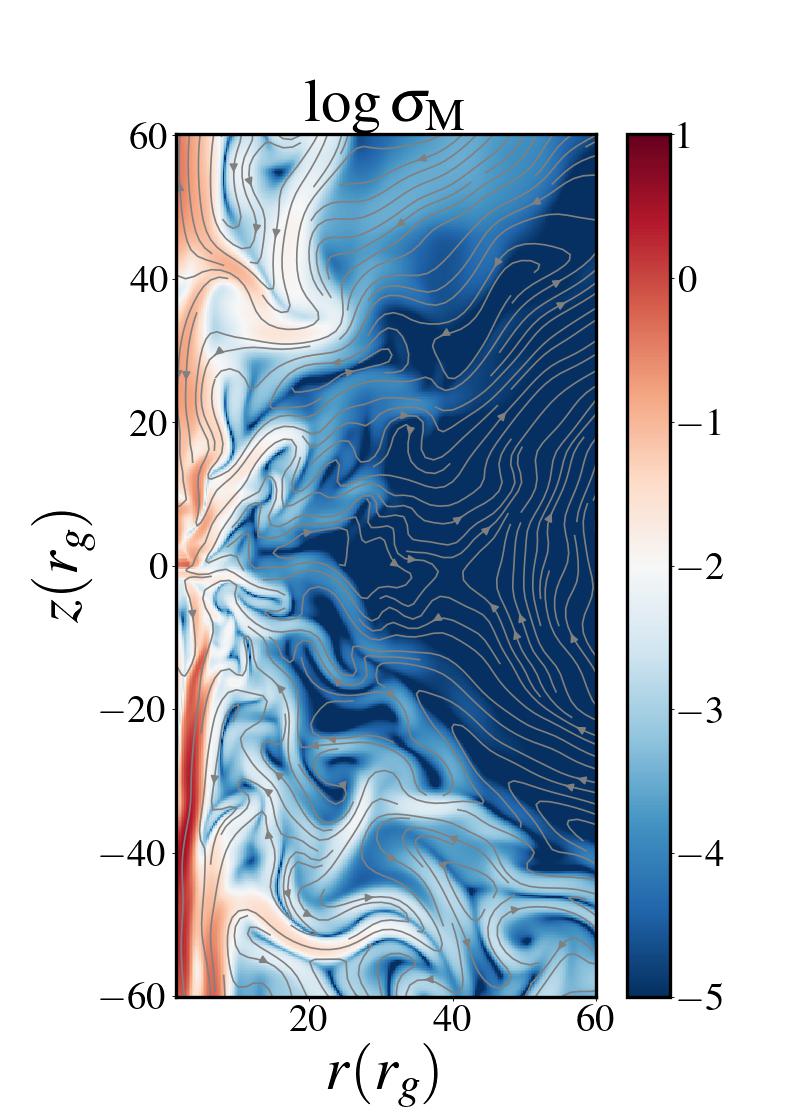} 
        \hskip -2mm
        \includegraphics[width=0.23\textwidth]{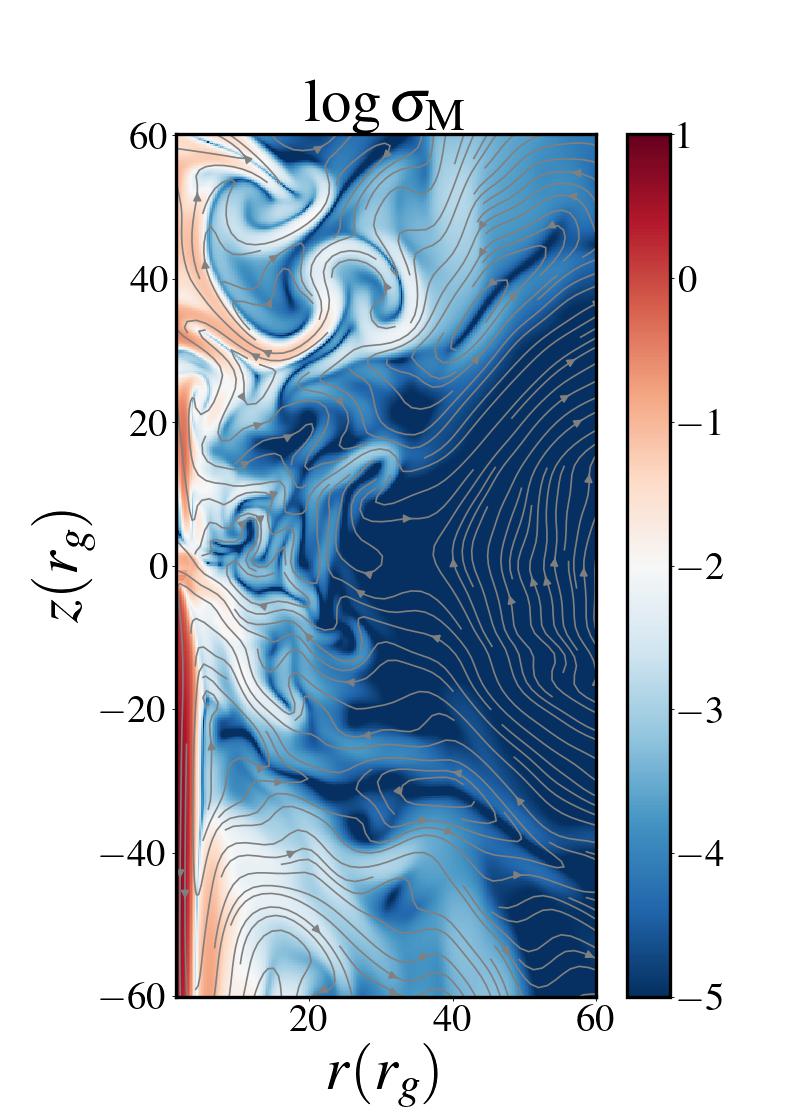} 
	\end{center}
	\caption{Distribution of gas density $(\rho)$, temperature $(T)$, plasma-$\beta$ $(\beta)$ and magnetization $(\sigma_{\rm M})$ for MAD state different black hole spin as $a_k = 0.99, 0.50, 0.0$ and $-0.99$. See the text for details.}
	\label{Figure_11}
\end{figure*}

\subsubsection{Luminosity and Spectral analysis: comparison of MAD and SANE state}
\label{spectral_ana}

In this section, we calculate the luminosity and spectral energy distribution (SED) for our model. In order to generate spectrum, we follow \citet{ Manmoto-etal-97} prescription to calculate radiative cooling processes. Here, we consider three radiative cooling processes, namely bremsstrahlung emission, synchrotron emission, and Comptonization of soft photons. Assuming a locally plane-parallel gas configuration at each radius ($r$) for the accretion flow, we calculate the radiation energy flux $F_\nu$ following \citet{Manmoto-etal-97} as
\begin{align} \label{spec_flux}
F_\nu = \frac{2 \pi}{\sqrt{3}} B_\nu \left[1 - \exp(-2 \sqrt{3} \tau_\nu^*)\right],
\end{align}
where $\tau_\nu^* = (\sqrt{\pi}/2) \kappa(0)H$ is the optical depth for absorption of the accretion in the vertical direction. Here $\kappa(0)$ is the the absorption coefficient on the equatorial plane.  The absorption coefficient $\kappa_{\nu}$ is defined as $\kappa_\nu = \chi_{\nu}/(4 \pi B_\nu)$. Here $\chi_\nu = \chi_{\nu,\rm br}+\chi_{\nu,\rm syn}$ is the total emissivity. 

Now, the bremsstrahlung emissivity $(\chi_{\nu, {\rm br}})$ can be written as \citep{Manmoto-etal-97}
\begin{align} \label{spec_brem_emisi}
\chi_{\nu, {\rm br}} = Q_{\rm br} \overline{G} \exp(\frac{h\nu}{k_{\rm B}T_e}),
\end{align}
where $Q_{\rm br}$ is the bremsstrahlung cooling rate as depicted in equation (\ref{Brem_rate}). Here $\overline{G}$ is the Gaunt factor, which is given by \citep{Rybicki-Lightman-79}
\[
   \overline{G} = 
\begin{cases}
    \frac{h}{k_{\rm B}T_e} \left(\frac{3}{\pi} \frac{k_{\rm B} T_e}{h \nu}\right)^{1/2}, & \text{if } \frac{k_{\rm B}T_e}{h \nu} < 1\\
    \frac{h}{k_{\rm B}T_e} \frac{\sqrt{3}}{\pi} \ln{\left(\frac{4}{\xi}\frac{k_{\rm B} T_e}{h \nu}\right)}, & \text{if } \frac{k_{\rm B}T_e}{h \nu} > 1.
\end{cases}
\]

We also consider the synchrotron emissivity $\chi_{\nu,\rm syn}$ by a relativistic Maxwellian distribution of electrons following \citet{Narayan-Yi95} as
\begin{align} \label{spec_sync_emisi}
\chi_{\nu,\rm syn} = 4.43 \times 10^{-30} \frac{4 \pi n_e \nu}{K_2 (1/\Theta_e)} I(x),
\end{align}
where $x = \frac{4 \pi m_e c \nu}{3 e B \Theta_e^2}$ and 
\begin{align}
I(x) = \frac{4.0505}{x^{1/6}} \left(1+\frac{0.40}{x^{1/4}}+ \frac{0.5316}{x^{1/2}}\right) \exp(-1.8899x^{1/3}).
\end{align}

Further, we consider the effect of Compton scattering. The energy enhancement factor $\eta$ due to the Compton scattering is defined as \citep{Manmoto-etal-97}
\begin{align} \label{spec_comp_factor}
\eta = e^{[s(A-1)]}[1-P(j_m+1, As)]+ \eta_{\rm max} P(j_m+1,s),
\end{align}
where $P$ is the incomplete gamma function and
\begin{align}
A = 1+4 \Theta_e+16\Theta_e^2,~~~ s = \tau_{\rm es} + \tau_{\rm es}^2,
\end{align}
where, $\eta_{\rm max} = 3 k_{\rm B}T_e/h \nu$ and $j_m = \ln{\eta_{\rm max}}/\ln{A}$. Also, $\tau_{\rm es}$ is the optical depth of the scattering as $\tau_{\rm es} = 2 n_e \sigma_{\rm T} H$ \citep{Esin-etal96}. Here, $\sigma_{\rm T}$ is the Thomson cross-section of the electron. 

Now, we evaluate the bremsstrahlung and synchrotron luminosity based on the frequency-dependent radiation flux following the equation (\ref{spec_flux}). The frequency-dependent bremsstrahlung ($L_{\rm br}$) and synchrotron ($L_{\rm syn}$) luminosity can be obtained throughout the disk surface as  
\begin{align}\label{brem_lumi}
L_{\rm br} = \int_S \int_\nu {F_{\nu},}_{\rm br}~d\nu~dS,
\end{align}
and
\begin{align}\label{sync_lumi}
L_{\rm syn} = \int_S \int_\nu {F_{\nu},}_{\rm syn}~d\nu~ dS,
\end{align}
where the surface integration is carried out all over the disk surface. $F_{\nu,\rm syn}$ and $F_{\nu,\rm br}$ can be calculated using synchrotron and bremsstrahlung emissivities using equation (\ref{spec_brem_emisi}) and (\ref{spec_sync_emisi}), respectively. On the other hand, the radiative luminosity can be calculated at the outer $z$-boundary and the outer $r$-boundary surfaces as follows 
\begin{align}\label{tot_lumi}
L_{\rm rad} = \int_S \bm{F}_{\rm rad}~ \bm{dS},
\end{align}
where $\bm{F}_{\rm rad}$ is the radiation energy ﬂux at the boundary surfaces.

 Here, we compare the bremsstrahlung, synchrotron, and radiative luminosity emanating from the whole computational regions for different magnetic fields $\beta_0=10, 25, 50$ and 100, respectively, depicted in Fig. \ref{Figure_6}a,b,c. We vary the frequency from $10^{10}$ to $10^{22}$ Hz to calculate frequency-dependent radiation flux. We observe that the bremsstrahlung luminosity is much higher compared to the synchrotron luminosity ($L_{\rm br} \gtrsim 10^{2}~L_{\rm syn}$), as shown in Figure. \ref{Figure_6}a,b. Because, in general, bremsstrahlung luminosity is proportional to $\rho^2$ and synchrotron luminosity is proportional to $\rho$. The bremsstrahlung luminosity is dominated due to the presence of highly dense thick torus in our model. We find that the value and nature of the bremsstrahlung luminosity are almost similar for different magnetic fields, shown in Fig. \ref{Figure_6} because the bremsstrahlung emissivity is independent of the magnetic field, as indicated in equation (\ref{spec_brem_emisi}). On the other hand, the synchrotron luminosity is comparatively higher for the MAD state compared to the SANE state due to the initial magnetic field strength, as depicted in Fig. \ref{Figure_6}b. We also show the radiative luminosity variation in Fig. \ref{Figure_6}c. We observe that the bremsstrahlung luminosities ($L_{\rm br} \sim 3 \times 10^{45}$) are almost the same as the radiative luminosities ($L_{\rm rad}$) towards the final stage of the simulation except for the weakest magnetic field case $\beta_0 = 100$. This is reasonable because free-free opacity is implicitly implemented in the PLUTO code, and the radiation field is solely based on only free-free emission. Moreover, the good agreement between $L_{\rm br}$ and $L_{\rm rad}$ shows that energy flux on the disk surface is well approximated in our torus problem by the diffusion approximation of the radiation vertically to the disk. Further, we analyze the power density spectrum (PDS) using bremsstrahlung luminosity ($L_{\rm br})$. In Fig. \ref{Figure_7}, we compare the PDS for MAD ($\beta_0 =10$) and SANE ($\beta_0 =100$) state. We observe peak frequency (fundamental) at around $\sim 1.52 \times 10^{-6}$ Hz for both MAD and SANE states using Lorentzian model. Thus, the radiative luminosity is found to vary quasi-periodically (QPOs) with periods of $\sim 8$ days. The QPOs are also observed for AGNs (e.g., Sgr A*) by other simulation studies \citep{Okuda-etal19, Okuda-etal22}. Interestingly, there is no clear distinction between MAD and SANE states as far as PDS of luminosity is concerned.

To generate the energy spectrum, we utilize equation (\ref{spec_flux}), (\ref{spec_brem_emisi}), (\ref{spec_sync_emisi}) and (\ref{spec_comp_factor}). Here, we compare spectral evolution with time for MAD ($\beta_0 = 10$) and SANE ($\beta_0 = 100$) state, as depicted in Fig. \ref{Figure_8}. The red, blue, green, and black curves are for simulation time $t=1000, 10000, 40000$, and $60000 t_g$, respectively. The solid and dotted curves for MAD and SANE state, respectively. With time, the synchrotron peaks become prominent in the SED. We observe two prominent peaks in each spectrum i.e., the synchrotron peaks at $\sim 10^{12-13}$ Hz and bremsstrahlung peaks at $\sim 10^{19-20}$Hz for MAD as well SANE state. We do not find any significant Compton enhancement in our model. The synchrotron peaks are $\gtrsim 5$ times brighter in magnitude for the MAD state compared to the SANE state but the bremsstrahlung peaks are similar for the MAD and SANE states. Moreover, we also observe that the characteristic nature of the spectrum is similar for MAD and SANE states as already explored by \citet{Xie-Zdziarski-19}. In general, the basic difference between MAD and SANE states is the strength of the magnetic field. But, the structure of the torus and radiation mechanism are similar for MAD and SANE states. Therefore, it is impossible to differentiate MAD to SANE state based on spectral analysis \citep{Xie-Zdziarski-19}. Further, we compare the spectrum by varying magnetic fields in Fig. \ref{Figure_9}. Here, we fix the magnetic field $\beta_0 = 10~(\rm{red}), 25~(\rm{blue}), 50~(\rm{green})$ and 100 (black) at the end of the simulation time $t=60000 t_g$. Here, we also observe that the nature of the spectrum is similar for all the cases and SED is brighter in synchrotron emission in the MAD state than SANE state.  

It is worth noting that the initial temperature distribution outside the torus in our model differs from typical GRMHD simulations. We assume that the atmosphere outside the torus consists of isothermal, nonrotating, high-temperature, and rarefied gas, as illustrated in Fig. \ref{Figure_01}b \citep{Kuwabara-etal05, Igarashi-etal20}. This differs significantly from the GRMHD model as implemented in the Black Hole Accretion Code (BHAC) \citep{Porth-etal-16}. In the BHAC code, the density and gas pressure outside the torus (atmosphere) are set to very small values: $\rho_{\rm atm} = 10^{-5} r^{-3/2}$ and $P_{\rm atm} = 10^{-7} r^{-5/2}$, respectively \citep{Porth-etal-16}. However, the outcome of our simulation results is consistent with the prescribed initial conditions of our model.


\subsection{MAD state: effect of black hole spin}
\label{effect_spin}
In this section, we investigate the effect of black hole spin on the highly magnetized accretion flow. For completeness, we examine the effect of black hole spin on the MAD state. Here we vary only the black hole spin ($a_k$) for fixed initial magnetic field $\beta_0 = 10$. We vary the black hole spin as $a_k = 0.99, 0.50, 0.0$ and -0.99. In Fig. \ref{Figure_10}a, we show the variation of mass accretion rate $(\dot{M}_{\rm acc})$ with time by varying the black hole spin $a_k$. The corresponding magnetic flux is shown in Fig. \ref{Figure_10}b. We observe that the accretion rate saturates near the Eddington rate ($\dot{M}_{\rm acc} \sim \dot{M}_{\rm Edd}$) in the MAD state for all the spin cases. Interestingly, we find the prograde flow reaches the MAD state much faster than the retrograde flow, as depicted in Fig. \ref{Figure_10}b. The MAD state is achieved at $t\sim 9500, 11000, 13000$ and $15000 t_g$  for $a_k =0.99, 0.50, 0.0$ and $-0.99$, respectively . Further, we present the distribution of density, temperature, plasma-$\beta$, and magnetization parameters in the first, second, third, and fourth row of Fig. \ref{Figure_11}, respectively. Here, we consider the simulation time at $t=40000 t_g$. It is observed that the black hole spin has minimal effect on the accretion process in the torus \citep{Hong-Xuan-etal23, Aktar-etal24}. Only the initial torus size slightly reduces for the lower spin values \citep{Aktar-etal24}. We also observe that MAD state is possible for prograde as well retrograde flow and the accretion properties are almost similar for all the spin values as depicted in Fig. \ref{Figure_11}. We find high magnetic pressure ($\beta << 1$) near the horizon for all spin values for the MAD state, shown in the third column of Fig. \ref{Figure_11}. Further, we observe high magnetization parameter $(\sigma_{\rm M} \sim 10)$ near the horizon for all the spin caese which is very common for the MAD state. The high magnetization region is the region for generating highly relativistic jets \citep{Dihingia-etal21, Dihingia-etal22, Hong-Xuan-etal23, Narayan-etal-22}.

\section{Summary and Discussions}
\label{conclusion}

In this work, we examine 2D relativistic, radiation, magnetohydrodynamics (Rad-RMHD) accretion flows around spinning AGN. To investigate the accretion flows, we consider an equilibrium torus around AGN following our previous work \citet{Aktar-etal24}. We consider the relativistic, radiation module in PLUTO code to model the accretion flow following \citet{Fuksman-Mignone-19}. The black hole gravity is modeled by effective Kerr potential \citep{Dihingia-etal18b} (see section \ref{Kerr_pot}). The initial magnetic field configuration is adopted following \citet{Hawley-etal02}, described in section \ref{Mag_config}. Our Rad-RMHD model is quite beneficial for exploring comparatively high spatial resolution and long-term simulation runs. In MHD flows, MRI gradually grows in the disk and Maxwell’s stress transports angular momentum outwards, leading to mass accretion \citep{Aktar-etal24}.  

We perform a comparative study, examining both MAD and SANE states. We observe MAD state for $\beta_0 = 10, 25$ and SANE state for $\beta_0 = 50, 100$ in our simulation model, shown in Fig. \ref{Figure_1}b. We examine the step-by-step accretion process in MAD state and the transition from SANE to MAD state in Fig. \ref{Figure_2}. The accumulation of a large amount of magnetic flux near the horizon during the transition from SANE to MAD state is a fundamental criterion for achieving the MAD state, as depicted in the plasma-$\beta$ plot in Fig. \ref{Figure_2} \citep{Narayan-etal-03, Igumenshchev-08, Tchekhovskoy-etal11, Narayan-etal12}. We also compare different flow variables for MAD and SANE states. We observe that the magnetization parameter is very high ($\sigma_{\rm M} \gtrsim 10$) near the horizon for the MAD state, while it remains very low ($\sigma_{\rm M} \lesssim 10^{-2}$) compared to the MAD state, as depicted in Fig. \ref{Figure_3}, second column. Moreover, we find that radiation energy density is very high for the MAD state compared to the SANE state near the horizon. Further, we observe that outward vertical velocity is close to relativistic near the funnel region for MAD compared to the SANE state, as shown in Fig. \ref{Figure_3} in the fourth column. 

In our simulation model, we adopt a single temperature radiation RMHD model in PLUTO code. However, we also employ a self-consistent two-temperature model to separately evaluate ion and electron temperatures based on our simulation model data, following the method described by \citet{Okuda-etal23}. Accordingly, we estimate synchrotron and bremsstrahlung luminosity considering the frequency-dependent radiation flux for various magnetic fields $\beta_0=10,25,50$ and 100, depicted in Fig. \ref{Figure_6}. The bremsstrahlung luminosity dominates over synchrotron due to the presence of a thick, dense torus in our model. We observe quasi-periodic oscillations (QPOs) behavior of the bremsstrahlung luminosity variation for all the magnetic field models. We find QPO peak frequency at $\sim 1.52 \times 10^{-6}$ Hz for MAD and SANE state, shown in Fig. \ref{Figure_7}. However, we do not find any characteristic difference in PDS for MAD and SANE states. In this regard, we also perform the time evolution and compare the spectral energy distribution (SED) for MAD and SANE state following \citet{Manmoto-etal-97}, shown in Fig. \ref{Figure_8}. We find that synchrotron emission is $5-10$ times brighter compared to SANE state but there is no difference in bremsstrahlung emission. Interestingly, it is difficult to distinguish the nature of the spectrum for MAD and SANE states by observation \citep{Xie-Zdziarski-19}. Finally, we investigate the effect of black hole spin on the MAD state, shown in Fig. \ref{Figure_10} and \ref{Figure_11}. It is observed that the MAD state is achieved for both prograde and retrograde accretion flow. 

The work uses a single temperature approximation, with the radiation transport mechanism considered to be only bremsstrahlung emission due to the difficulty in specifying frequency-independent opacity for the synchrotron emission. However, it is important to investigate a proper two-temperature model by including other inevitable cooling mechanisms such as synchrotron and Comptonization in the PLUTO code \citep{Ryan-etal-18, Dihingia-etal22, Dihingia-etal23, Hong-Xuan-etal23}. This will help in accurately understanding the spectral evolution for MAD and SANE states. To gain a complete understanding of the global behavior of highly magnetized flow, a 3D simulation model is necessary. We plan to examine a 3D global radiation simulation in the future.


\section*{Acknowledgments}
We express our sincere gratitude to the anonymous referee for the valuable suggestions and comments that have contributed to the improvement of the manuscript. This work is supported by the National Science and Technology Council of Taiwan through grant NSTC 112-2811-M-007-038 and 112-2112-M-007-040, and by the Center for Informatics and Computation in Astronomy (CICA) at National Tsing Hua University through a grant from the Ministry of Education of Taiwan. The simulations and data analysis have been carried out on the CICA Cluster at National Tsing Hua University. 






\bibliography{refs}{}
\bibliographystyle{aasjournal}


\end{document}